\documentclass{fundam}
\pdfoutput=1
\usepackage{amssymb}
\usepackage{url}
\usepackage{graphics}
\usepackage{graphicx}
\usepackage{fancyvrb}
\usepackage{epsfig}
\usepackage{amsmath}
\usepackage{float}
\usepackage{epstopdf}
\usepackage{subfigure}

\newtheorem{Lemma}{Lemma}
\usepackage{color}
\graphicspath{{fig/}}

\begin{document}


\setcounter{page}{11}
\publyear{24}
\papernumber{2102}
\volume{193}
\issue{1}

  \versionForARXIV


\title{On Completely Edge-Independent Spanning Trees \\ in Locally Twisted Cubes}

\author{Xiaorui Li\thanks{Also works: Provincial Key Laboratory for Computer Information Processing Technology,
          Soochow University, Suzhou 215006, China. \newline
          $^c$Address for correspondence: School of Computer Science and Technology, Soochow
          University, Suzhou 215006, China. \newline \newline
          \vspace*{-6mm}{\scriptsize{Received January 2024; \ accepted November 2024.}}},   Baolei Cheng$^{c,*}$,  Jianxi Fan, Yan Wang
                                     \\
School of Computer Science and Technology, Soochow University,\\
Suzhou 215006, China\\
20224227038@stu.suda.edu.cn,\{chengbaolei,jxfan,wangyanme\}@suda.edu.cn
\and
 Dajin Wang
 \\
School of Computing, Montclair State University\\
Upper Montclair, NJ 07043, USA \\
wangd@montclair.edu
}

\maketitle

    \runninghead{X. Li et al.}{On Completely Edge-Independent Spanning Trees in Locally Twisted Cubes}

\begin{abstract}
\noindent
A network can contain numerous spanning trees. If two spanning trees $T_i,T_j$ do not share any common edges, $T_i$ and $T_j$ are said to be {\it pairwisely edge-disjoint}. For spanning trees $T_1, T_2, ..., T_m$, if every two of them are pairwisely edge-disjoint, they are called {\it completely edge-independent spanning trees} (CEISTs for short). CEISTs can facilitate many network functionalities, and constructing CEISTs as maximally allowed as possible in a given network is a worthy undertaking.
In this paper, we establish the maximal number of CEISTs in the {\it locally twisted cube} network, and propose an algorithm to construct $\lfloor \frac{n}{2} \rfloor$ CEISTs in $LTQ_n$, the $n$-dimensional locally twisted cube. The proposed algorithm has been actually implemented, and we present the outputs. Network broadcasting in the $LTQ_n$ was simulated using $\lfloor\frac{n}{2}\rfloor$ CEISTs, and the performance compared with broadcasting using a single tree.

\medskip\noindent
\textbf{Keywords:} Broadcasting; Edge-disjoint; CEISTs; Locally twisted cubes; Spanning trees; Tree embedding.
\end{abstract}

\section{Introduction}

 One important indicator for a network's robustness is its ability to effectively embed useful structures, such as paths, rings, trees, etc. that satisfy certain constraints in the network. In recent years, the question of finding {\em completely edge-independent spanning trees} (CEISTs) in interconnection networks has particularly interested researchers in the field.
The completely edge-independent spanning trees in a network can be an instrumental facilitator of many network functions, such as reliable communication, fault-tolerant broadcasting, secure messages distribution, etc [1, 2, 3, 4]. If we can determine the maximum number of CEISTs to optimize the utilization of the edges in a given network, we can achieve the maximum channel utilization, and reduce the communication delay. If there are $k$ CEISTs in a network, then the network can tolerate, in the worst-case scenario, as many as $(k-1)\times (n-1)$ tree-edges to be faulty, and still achieve broadcasting.

In the study of computer networks, a network is almost exclusively modeled by a {\it graph} $G=(V,E)$ as in graph theory, with $V$ being the set of vertices (nodes in network terms), and $E$ being the set of edges (links/lines/channels between nodes). In this paper, these terms---graphs and networks, vertices and nodes, edges and links/lines/channels---will be used interchangeably.

There are several different categories of independent spanning trees, and they are broadly classified into {\em edge-independent spanning trees} (EISTs), {\em node-independent spanning trees} (NISTs), {\em completely independent spanning trees} (CISTs), and {\em edge-disjoint spanning trees} (EDSTs), etc. (see [4] for all these definitions). In general, EISTs is a set of spanning trees rooted at $u$ in $G$ such that there are no common internal edges between $u$ and any other node among the paths in these spanning trees [4], and the difference between EISTs and EDSTs is that in EISTs, all spanning trees share one particular node as their common root, while all spanning trees in EDSTs are rootless. For that reason, CEISTs as EDSTs have been interchangeably used in the literature.

However, we noted that the definition of EDSTs in [21] is actually our definition of EISTs. Furthermore, Lin et al. proved that Hsieh and Tu’s spanning trees are indeed node-independent spanning trees [27], that is, the spanning trees they constructed are rooted trees. Therefore, the spanning trees dealt with in our paper are different than those in [21]. Following the nomenclature of CIST (where “completely” means “rootless”), in the rest of this paper we employ the term CEISTs instead of EDSTs to avoid further confusion.

The {\it hypercube}, denoted by $Q_n$, is a classical interconnection network that has many desirable properties, such as low diameter, high connectivity, symmetry, etc. [5, 6, 7]. The CEISTs in $Q_n$ have been extensively studied.
Barden et al. proposed a method for obtaining the maximum number of CEISTs in $Q_n$, and proved that there exist $\lfloor \frac{n}{2} \rfloor$ CEISTs in $Q_n$ [8]. For the {\it Cartesian product network} $G\times F$, Ku et al. gave two methods to embed CEISTs [9]. The first method constructed $n_1+n_2$ CEISTs in $G \times F$ with certain assumptions, while the second one with no assumptions constructed $n_1+n_2-1$ CEISTs in $G \times F$, where $n_1$ (resp. $n_2$) is the number of CEISTs in $G$ (resp. $F$). In 2021, Zhou et al. proposed the maximum number of CEISTs and edge-disjoint spanning $c$-forests of equiarboreal graphs [10]. In 2022, Wang et al. proposed an algorithm that constructed completely independent spanning trees in the line graph of graph $G$ based on $G$'s CEISTs [15]. Furthermore, CEISTs have also been studied for some special networks [11, 12, 13, 14].

\eject
Fan et al. introduced the notion of {\it bijective connection networks} (BC networks for short) [16, 28], which include many well-known networks, such as hypercubes, locally twisted cubes, crossed cubes, etc., and other less known networks. Based on BC networks, {\it conditional BC networks} were defined by Cheng et al. in [17]. It has been shown that hypercubes, crossed cubes, locally twisted cubes, and M\"obius cubes all belong to conditional BC networks. As of today, CEISTs in hypercubes [8] and crossed cubes [18] have been obtained, while CEISTs in locally twisted cubes and M\"obius cubes [24] still remain unsolved.

\medskip
As a prominent variant of the hypercube, the {\it locally twisted cube} $LTQ_n$, proposed by Yang et al. [19], has advantageous properties including low diameter and low fault-diameter. Hsieh et al. proved that $LTQ_n$ is $(2n-5)$-Hamiltonian (for $n \geq 3$) if each vertex in $LTQ_n$ is associated with at least two fault edges [20].
In this paper, we study how to embed the maximal number of CEISTs in the $LTQ_n$. Our work in this paper can be outlined as follows:
\begin{enumerate}
\item[1)] We propose a recursive algorithm, named {\it CEISTs\_$LTQ$}, that constructs $\lfloor \frac{n}{2} \rfloor$ CEISTs $T_1, T_2, ...,\newline T_{\lfloor \frac{n}{2} \rfloor}$ in the $LTQ_n$;
\item[2)] We prove theoretically the correctness of {\it CEISTs\_$LTQ$}, and determine its time complexity as $O(n\cdot{2^n})$, where $n$ is the dimension of $LTQ$;
\item[3)] To solidly validate the algorithm, {\it CEISTs\_$LTQ$} has been actually implemented, and we present the running outputs;
\item[4)] We have simulated the broadcasting in $LTQ_n$ using {\it CEISTs}, and the outcomes are presented and discussed.
\end{enumerate}

It is worth pointing out that the CEISTs-construction methods developed earlier for the hypercube and the crossed cube [8, 18] cannot be directly applied to {\it LTQ}. The link connection between subcubes in {\it LTQ} is more complicated than in the hypercube/crossed cube, calling for different, new approaches in the CEISTs-construction process.

\medskip
The rest of this paper proceeds as follows. Section 2 provides the preliminaries. Section 3 includes the paper's main work---describing the algorithm for constructing $\lfloor \frac{n}{2} \rfloor$ CEISTs in $LTQ_n$, proving its correctness, and analyzing its time complexity. Section 3 also discusses the technical similarity/difference between {\it CEISTs\_$LTQ$} and existing CEISTs algorithms for hypercubes and crossed cubes. Section 4 presents the simulation experiments to verify the validity of the algorithm, and evaluates its performance in efficient broadcasting. Section 5 concludes the paper.

\section{Preliminaries}

\subsection{Terminology and notation}

 A network can be abstracted as a graph $G(V(G),E(G))$, where $V(G)$ denotes the vertex set and $E(G)$ denotes the edge set, representing the servers and the links between them respectively. As pointed out earlier, graphs and networks are interchangeably used throughout this paper.

$T$ is said to be a spanning tree of graph $G$, if (1) $T$ contains all the vertices of $G$; (2) $T$ has exactly $|V(G)| - 1$ edges from $E(G)$. A path started from $u$ and ended at $v$ is denoted $(u, v)$-path. Given two $(u, v)$-paths $P$ and $Q$ started at $u$ and ended with $v$, $P$ and $Q$ are \emph{edge-disjoint} if they share no common edges. Let $T_1,T_2,...,T_m$ be spanning trees in graph $G$, if for every pair of them do not contain the common edges, then $T_1,T_2,...,T_m$ are called completely edge-independent spanning trees (CEISTs for short) in $G$.

A binary string $u$ with length $n$ can be written as $u_{n-1} u_{n-2} ...u_i u_{i-1} ...u_0$ ,where $u_i\in \left\{0,1\right\}$ and $0 \leq i \leq n-1$. The complement of $u_i$ will be denoted by $\overline{u_i}$ ($\overline{0} = 1$  and $\overline{1} = 0) $. $|u|$ is the decimal value of $u$. A path $P$ from $v^{(1)}$ to $v^{(n)}$ can be denoted as $v^{(1)}$-$v^{(2)}$-$\cdots$-$v^{(n)}$. $|u| - |v|$ means the decimal value of vertex $u$ minus the decimal value of vertex $v$.

\subsection{Locally twisted cubes}

 We adopt the definition of $LTQ_n$ as follows.

\medskip
\noindent {\bf Definition 1} [19].
For integer $n \geq 2$. The $n$-dimensional locally twisted cube, denoted by $LTQ_n$, is defined recursively as follows.

(1) $LTQ_2$ is a graph consisting of four vertices labeled with $00, 01, 10$, and $11$, respectively, connected by four edges $(00,01), (00,10), (01,11)$, and $(10,11)$.

(2) For integer $n \geq 3$, $LTQ_n$ is built from two disjoint copies of $LTQ_{n-1}$ according to the following steps. Let $LTQ_{n-1}^{0}$ (respectively, $LTQ_{n-1}^{1}$) denote the graph obtained by prefixing the label of each vertex in one copy of $LTQ_{n-1}$ with 0 (respectively, 1). Each vertex $u = 0(u_{n-2}u_{n-3}...u_0)$ in $LTQ_{n-1}^{0}$ is connected with the vertex $1(u_{n-2} \oplus u_0)  u_{n-3}...u_0$ in $LTQ_{n-1}^{1}$ by an edge, where“$\oplus$”represents the modulo 2 addition.

\begin{figure}[!b]
\vspace*{-2mm}
\centering
\subfigure[]{
\label{LTQ3}
\includegraphics[height=2.9cm]{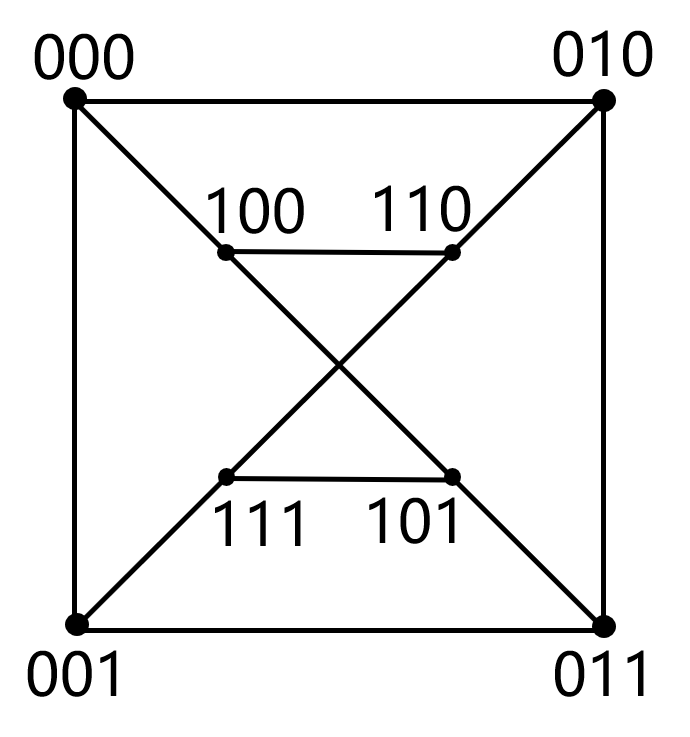}}
\subfigure[]{
\label{LTQ4}
\includegraphics[height=3.4cm]{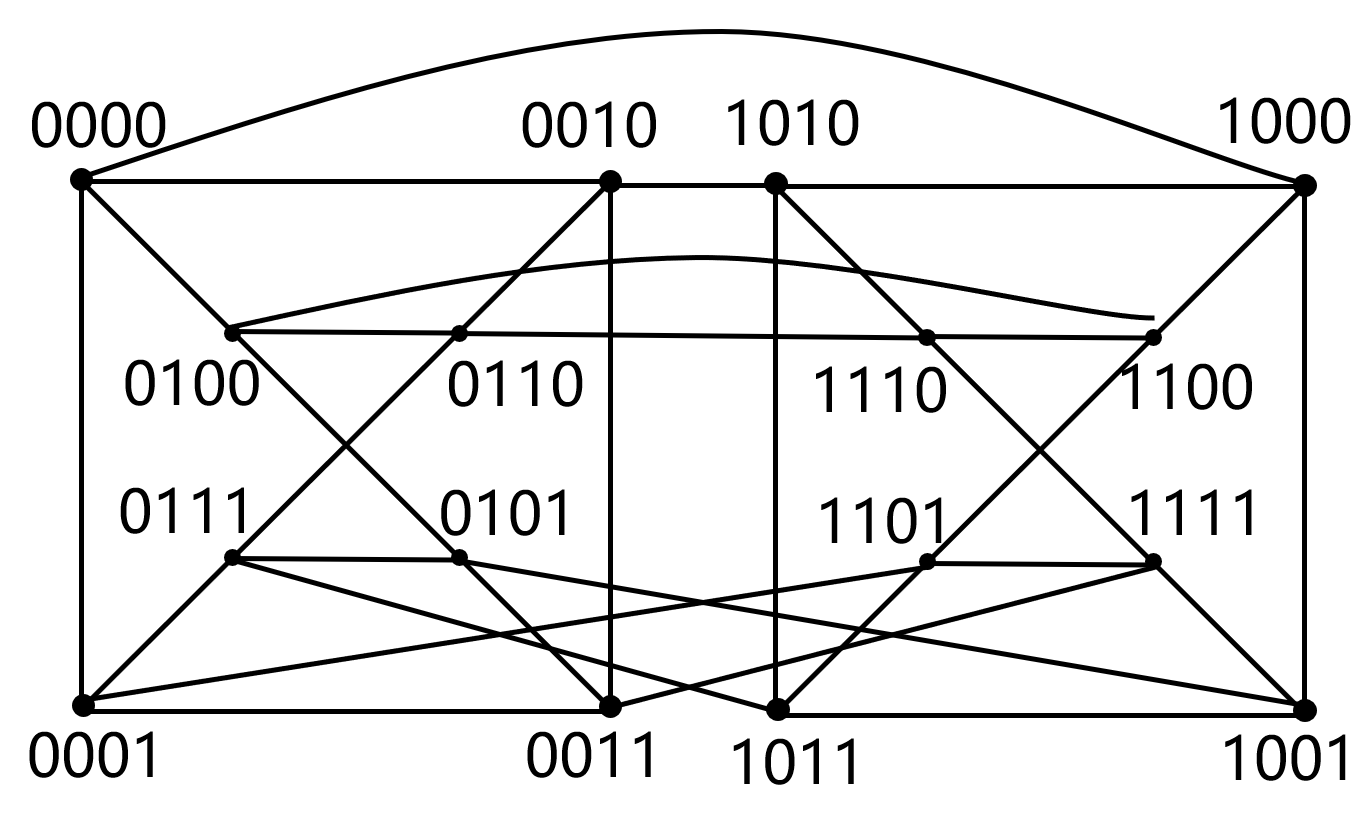}}\vspace*{-1mm}
\caption{(a) $LTQ_3$;\, (b) $LTQ_4$.}
\label{fig1}\vspace*{-1mm}
\end{figure}

\smallskip
The locally twisted cube $LTQ_n$ can also be equivalently defined in the following non-recursive fashion.

\medskip
\noindent {\bf Definition 2} [19]. For integer $n \geq 2$, the $n$-dimensional locally twisted cube, $LTQ_n$, is a graph with $\left\{0,1\right\}^n$ as the vertex set. Two vertices $u = u_{n-1}u_{n-2}...u_0$ and $v = v_{n-1}v_{n-2}...v_0$ in $LTQ_n$ are adjacent if and only if either (1) For some $2 \leq i \leq n-1$, $u_i = \overline{v_i}$ and $u_{i-1} = v_{i-1}\oplus u_0$ , and $u_j = v_j$ for all the remaining bits or, (2) For some $i \in \left\{0,1\right\}$, $u_i = \overline{v_i}$ , and $u_j = v_j$ for all the remaining bits.

\smallskip
Fig. \ref{fig1} shows the examples of $LTQ_3$ and $LTQ_4$.

\medskip
It follows that $LTQ_n$ is an $n$-regular graph, and any two adjacent vertices in $LTQ_n$ differ by at most two consecutive bits. Moreover, the following lemma holds.

\begin{Lemma}\label{lem1} \rm[21] Let $u = u_{n-1}u_{n-2}...u_0$ and $v = v_{n-1}v_{n-2}...v_0$ be two adjacent vertices in $LTQ_n ( n \geq 2)$ with $u > v$. Then, the following statements hold.

(1) If $|u|$ is even, then  $|u| - |v|$ = $2^i$ for some $0 \leq i \leq {n-1}$.

(2) If  $|u|$ is odd, then either $|u| - |v|$ = $2^i$ for some $i \in \left\{0,1\right\}$ or $|u| - |v|$ = $2^i - \left[ (-1)^{u_{i}-1} \times 2^{i-1} \right]$ for some $i \geq 2$.
\end{Lemma}

When two adjacent vertices $u$ and $v$ have a leftmost differing bit at position $d$, that is $u_{d} \neq v_{d}$ and $u_{n-1}u_{n-2}...u_{d+1} = v_{n-1}v_{n-2}...v_{d+1}$. Furthermore, for $W \subseteq V(LTQ_n), N_d(W) = \left\{ N_d(w)|w \in W \right\}$ denotes the set of the $d$-neighbors of all vertices in $W$. When two vertices $u$ and $v$ are adjacent, and both $u$'s decimal value and $v$'s decimal value are even (odd), we say $(u,v)$ is even (odd) edge.

\section{Completely edge-independent spanning trees in locally twisted cubes}

     In this section, we propose an algorithm to construct $\lfloor \frac{n}{2} \rfloor$ CEISTs in $LTQ_n$, where $n \geq 2$, prove its correctness, and analyze its time complexity. In what follows, we represent each vertex in the trees by its decimal value.

\subsection{Construction Algorithm of CEISTs for locally twisted cubes}

\begin{figure}[htbp]
\centering
\includegraphics[height=2.3cm]{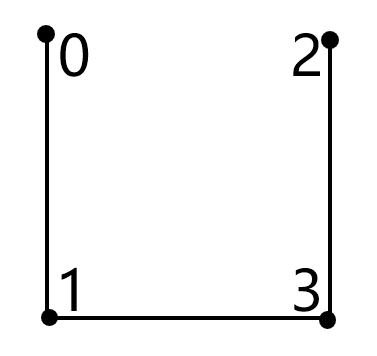}
\caption{CEIST $T_1$ in $LTQ_2$.}
\label{fig2}
\end{figure}

 A CEIST can be constructed easily in $LTQ_2$ and an example is presented in Fig. \ref{fig2}.

For $n \geq 3$ and $n$ is odd, $LTQ_n$ consists of two subcubes $LTQ_{n-1}^0$ and $LTQ_{n-1}^1$, denoted by $A$ and $B$. If there are $\lfloor \frac{n}{2} \rfloor$ CEISTs in $A$, we can construct $\lfloor \frac{n}{2} \rfloor$ CEISTs
that are one-to-one isomorphic with $\lfloor \frac{n}{2} \rfloor$ trees of $A$ in $B$ accordingly. Then, for every two isomorphic trees in $LTQ_n$, we expect to connect them through specific edges to obtain $\lfloor \frac{n}{2} \rfloor$ CEISTs in $LTQ_n$. Based on the above discussion, we propose a function (see algorithm {\it Odd\_CEISTs}) to obtain $\lfloor \frac{n}{2} \rfloor$ CEISTs in $LTQ_n$.

\begin{Verbatim}[commandchars=\\\{\},
framerule=0.6mm,frame=lines,codes={\catcode`$=3\catcode`^=7\catcode`_=8}]
\textbf{Algorithm {\it Odd\_CEISTs}}
\textbf{Input:} $\gamma_A (n-1)$, and $P_A: v_A^{1}$-$v_A^{2}$-$\cdots$-$v_A^{\lceil\frac{n}{2}\rceil}$.
($P_A$ is constructed recursively by algorithm{\it CEISTs\_$LTQ$}).
\textbf{Output:} $\gamma (n)$.
\textbf{Begin}
\textbf{Step 1.} Construct $T_1^B$, $T_2^B$,..., $T_{\lfloor\frac{n}{2}\rfloor}^B$ and the path $P_B: v_B^{1}$-$v_B^{2}$-$\cdots$-$v_B^{\lceil\frac{n}{2}\rceil}$.
\hspace{5mm}1: \textbf{for} $i=1$ to $\lfloor\frac{n}{2}\rfloor$ \textbf{do}
\hspace{5mm}2:   Construct tree $T_i^B$ by adding $2^{n-1}$ to each vertex in $T_i^A$.
\hspace{5mm}3: \textbf{end for}
\hspace{5mm}4: \textbf{for} $i=1$ to $\lceil\frac{n}{2}\rceil$ \textbf{do}
\hspace{5mm}5:   $|v_B^{i}| = |v_A^{i}| + 2^{n-1}$.
\hspace{5mm}6: \textbf{end for}
\textbf{Step 2.} Construct $T_1$, $T_2$,..., and $T_{\lfloor\frac{n}{2}\rfloor}$ as follows.
\hspace{5mm}7: \textbf{for} $i=1$ to $\lfloor\frac{n}{2}\rfloor$ \textbf{do}
\hspace{5mm}8:   $V(T_i) = V(LTQ_n)$.
\hspace{5mm}9:   $E(T_i) = E(T_i^A)\cup{E(T_i^B)}\cup{\{(v_A^i,v_B^i)\}}$.
\hspace{5mm}10: \textbf{end for}
\hspace{5mm}11: return $\gamma (n)$ = \{$T_1$, $T_2$,..., $T_{\lfloor\frac{n}{2}\rfloor}$\}.
end
\end{Verbatim}

\begin{figure}[htb]
\vspace*{-2mm}
\centering
\includegraphics[height=3cm]{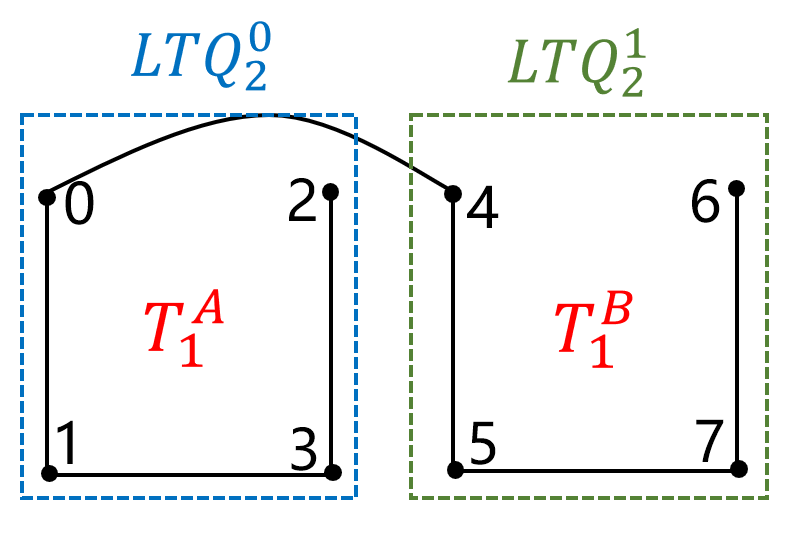}
\caption{CEIST $T_1$ in $LTQ_3$.}
\label{fig3}
\end{figure}

\begin{figure}[htb]
\centering
\subfigure[$T_1$]{
\label{LTQ5-T1}
\includegraphics[height=5cm]{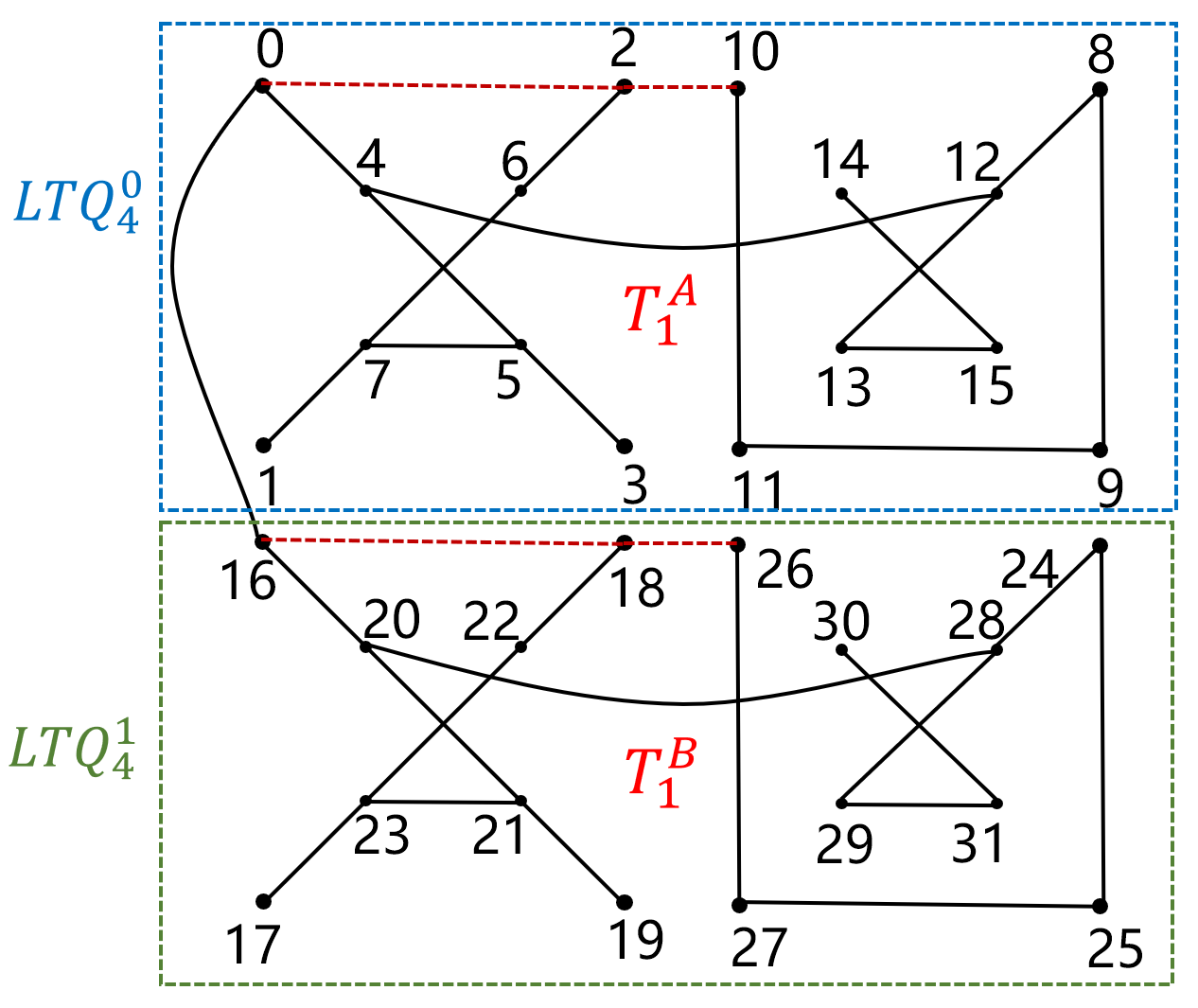}}
\subfigure[$T_2$]{
\label{LTQ5-T2}
\includegraphics[height=5cm]{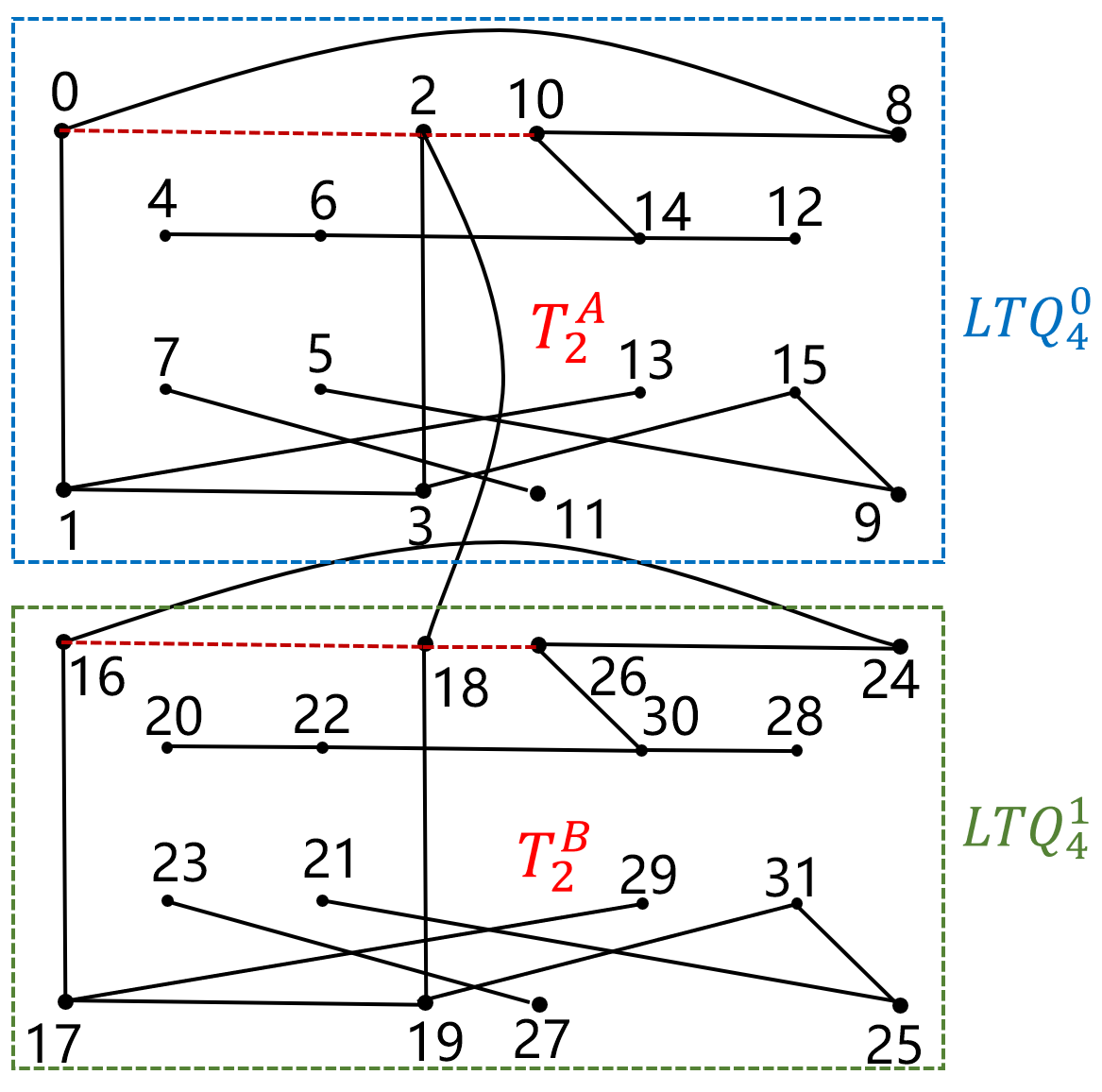}}
\caption{Two CEISTs $T_1$ and $T_2$ in $LTQ_5$.}
\label{fig4}\vspace*{-2mm}
\end{figure}

\medskip
\noindent \textbf{Example 1.} By algorithm {\it Odd\_CEISTs}, Fig. \ref{fig3} demonstrates the construction of CEIST $T_1$ in $LTQ_3$, and
Fig. \ref{fig4} demonstrates the construction of two CEISTs $T_1$ and $T_2$ in $LTQ_5$. We take the $LTQ_5$ as an example. Firstly, the two CEISTs $\gamma_A (4)= \{T_1^A, T_2^A\}$ and the path $P_A: 0$-$2$-$10$ in $LTQ_4$ as the input of the algorithm are presented in Fig.~5. Next, according to the step 1 of the algorithm, $T_1^B$, $T_2^B$ and $P_B: 16$-$18$-$26$ are obtained. Then, according to the step 2 of the algorithm, $T_1$ is constructed by connecting $T_1^A$ and $T_1^B$ through edge (0, 16). $T_2$ is constructed by connecting $T_2^A$ and $T_2^B$ through edge (2, 18).

For $n \geq 4$ and $n$ is even, $LTQ_n$ consists of four subcubes $LTQ_{n-2}^{00}$, $LTQ_{n-2}^{10}$, $LTQ_{n-2}^{11}$ and $LTQ_{n-2}^{01}$, denoted by $A$, $B$, $C$ and $D$, which have the common prefix 00, 10, 11 and 01, respectively. If there are $\frac{n-2}{2}$ CEISTs in $A$, we can construct $\frac{n-2}{2}$ CEISTs that are one-to-one isomorphic with $\frac{n-2}{2}$ trees of $A$ in $B$ accordingly, and so are $C$ and $D$. Then, for every four isomorphic trees in $LTQ_n$, we expect to connect them through specific edges to obtain $\lfloor \frac{n}{2} \rfloor$ CEISTs in $LTQ_n$, and after the construction of $\lfloor \frac{n}{2} \rfloor$ CEISTs, we also expect to get $\frac{n}{2}$ unused edges to form exactly path. Based on the above discussion, we propose a function (see algorithm {\it Even\_CEISTs}) to obtain $\lfloor \frac{n}{2} \rfloor$ CEISTs in $LTQ_n$.
\medskip

\begin{Verbatim}[commandchars=\\\{\},
framerule=0.6mm,frame=lines,codes={\catcode`$=3\catcode`^=7\catcode`_=8}]
\textbf{Algorithm {\it Even\_CEISTs}}
\textbf{Input:} $\gamma_A (n-2)$ and $P_A: v_A^{1}$-$v_A^{2}$-$\cdots$-$v_A^{\frac{n}{2}}$.
($P_A$ is constructed recursively by algorithm{\it CEISTs\_$LTQ$}).
\textbf{Output:} $\gamma (n)$, and $P: v^{1}$-$v^{2}$-$\cdots$-$v^{\frac{n}{2}+1}$.
\textbf{Begin}
\textbf{Step 1.} For $1 \leq i \leq \frac{n-2}{2}$, construct $T_i^B, T_i^C, T_i^D, P_B, P_C, P_D$.
\hspace{5mm}1: \textbf{for} $i=1$ to $\frac{n-2}{2}$ \textbf{do}
\hspace{5mm}2:   Construct $T_i^B, T_i^C, T_i^D$ by adding $2^{n-1}, 3*2^{n-2}, 2^{n-2}$ to
\hspace{5mm}each vertex in $T_i^A$, respectively.
\hspace{5mm}3: \textbf{end for}
\hspace{5mm}4: Construct paths $P_B, P_C, P_D$ by adding $2^{n-1}, 3*2^{n-2}, 2^{n-2}$ to
\hspace{5mm}each vertex in $P_A$, respectively.
\textbf{Step 2.} Construct $T_i$, for $1 \leq i \leq \frac{n}{2}-2$.
\hspace{5mm}5:   $V(T_i) = V(LTQ_n)$,
\hspace{5mm}6:   $E(T_i) = E(T_i^{\epsilon\in\{A,B,C,D\}})\cup\{(v_A^i,v_B^i)\}\cup\{(v_B^i,v_C^i)\}\cup\{(v_C^i,v_D^i)\}$.
\textbf{Step 3.} Construct $T_{\frac{n}{2}-1}$($i = \frac{n}{2}-1$).
\hspace{5mm}7: $V(T_i) = V(LTQ_n)$,
\hspace{5mm}8: $E(T_i) = E(T_i^{\epsilon\in\{B,C,D\}})\cup\{(v_B^i,v_C^i)\}\cup\{(v_C^i,v_D^i)\}$
\hspace{5mm}            $\cup\{(v, N_{n-2}(v))|v \in V(T_i^A)\}$.
\textbf{Step 4.} Construct $T_{\frac{n}{2}}$($i = \frac{n}{2}$).
\hspace{5mm}9: $V(T_i) = V(LTQ_n)$,
\hspace{5mm}10: $E(T_i) = E(T_{i-1}^A) \cup \{(v, N_{n-1}(v))| v \in V(T_{i-1}^A) and |v| is even\}$
              $\backslash \{(v_A^j,v_B^j) | j = 1 to i-2, and j = i\}$
              $\cup \{(v, N_{n-2}(v))| v \in V(T_{i-1}^B)and |v| is even\}$
              $\backslash \{(v_B^j,v_C^j) | j = 1 to i-1\}$
              $\cup \{(v, N_{n-1}(v))| v \in V(T_{i-1}^C) and |v| is even\}$
              $\backslash \{(v_C^j,v_D^j) | j = 1 to i-1\}$
              $\cup \{(v, N_{n-1}(v)) | v \in V(T_{i-1}^A) and |v| is odd\}$
              $\cup \{(v, N_{n-2}(v)) | v \in V(T_{i-1}^C) and |v| is odd\}$
              $\cup \{(v, N_{n-1}(v)) | v \in V(T_{i-1}^B) and |v| is odd\}$
              $\cup \{(v_r^j, v_r^{j+1}) | r \in \{B, C, D\}, j = 1 to i-1\}$.
\textbf{Step 5.} Construct the path $P: v^{1}$-$v^{2}$-$\cdots$-$v^{\frac{n}{2}+1}$ as follows.
\hspace{5mm}11: $v^{i} = v_A^{i}$, for $1 \leq i \leq \frac{n}{2}$.
\hspace{5mm}12: $v_A^{\frac{n}{2}+1} = v_A^{\frac{n}{2}} + 2^{n-1}$.
\hspace{5mm}13: return $\gamma (n)$ = \{$T_1$, $T_2$,..., $T_{\frac{n}{2}}$\}, and $P$.
end
\end{Verbatim}

\begin{figure}[!h]
\centering
\subfigure[$T_1$]{
\label{LTQ4-T1}
\includegraphics[height=5cm]{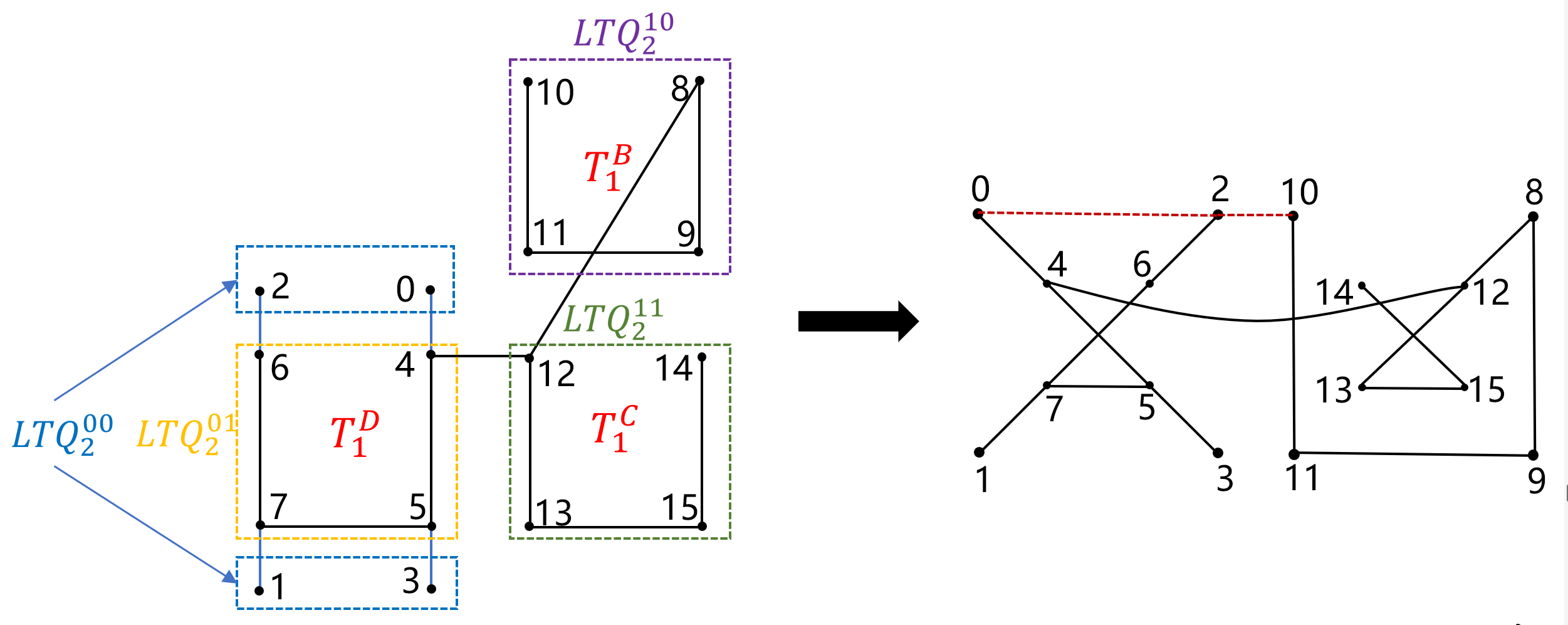}}
\subfigure[$T_2$]{
\label{LTQ4-T2}
\includegraphics[height=5cm]{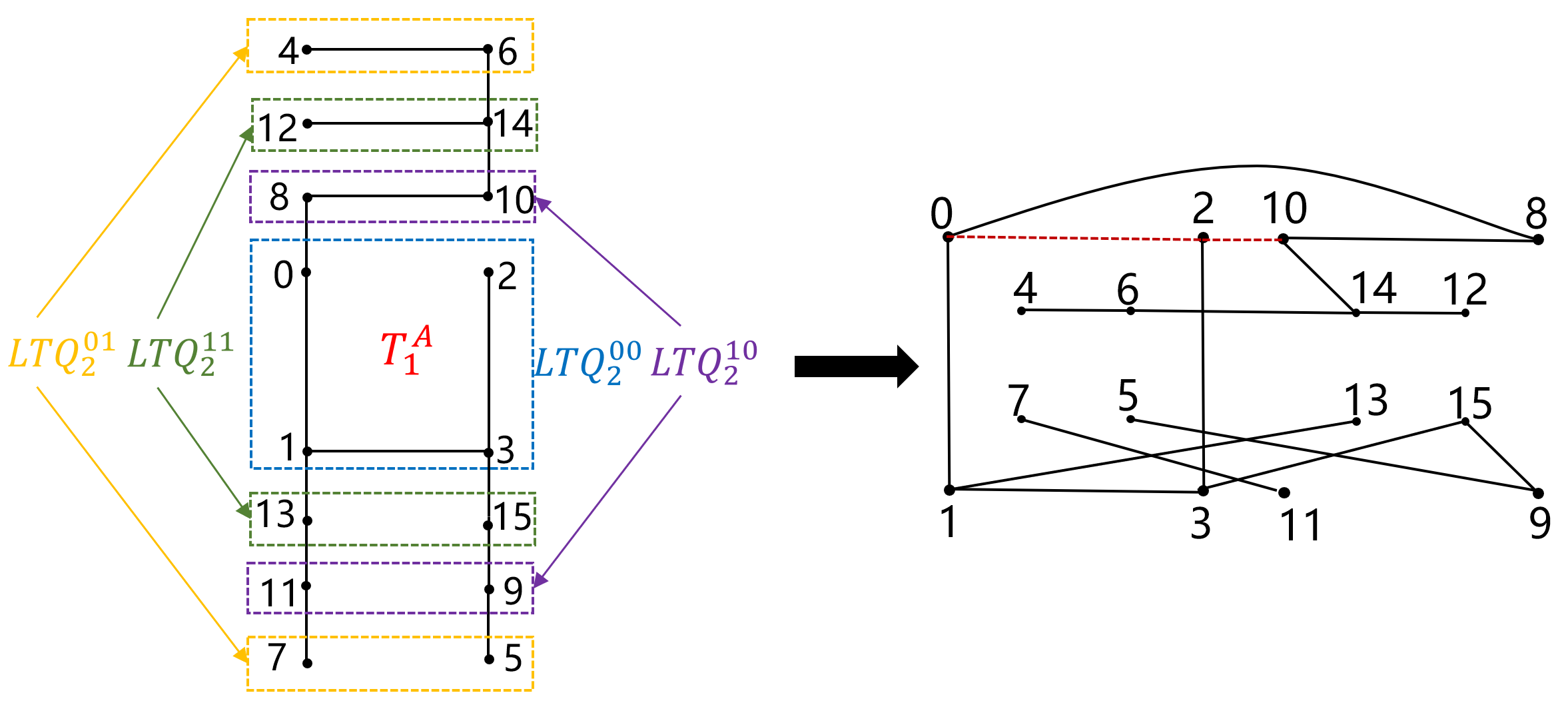}}
\caption{Two CEISTs $T_1$ and $T_2$ in $LTQ_4$.}
\label{fig5}
\end{figure}

\begin{figure}[!h]
\centering
\subfigure[$T_1$]{
\label{LTQ6-T1}
\includegraphics[height=5.5cm]{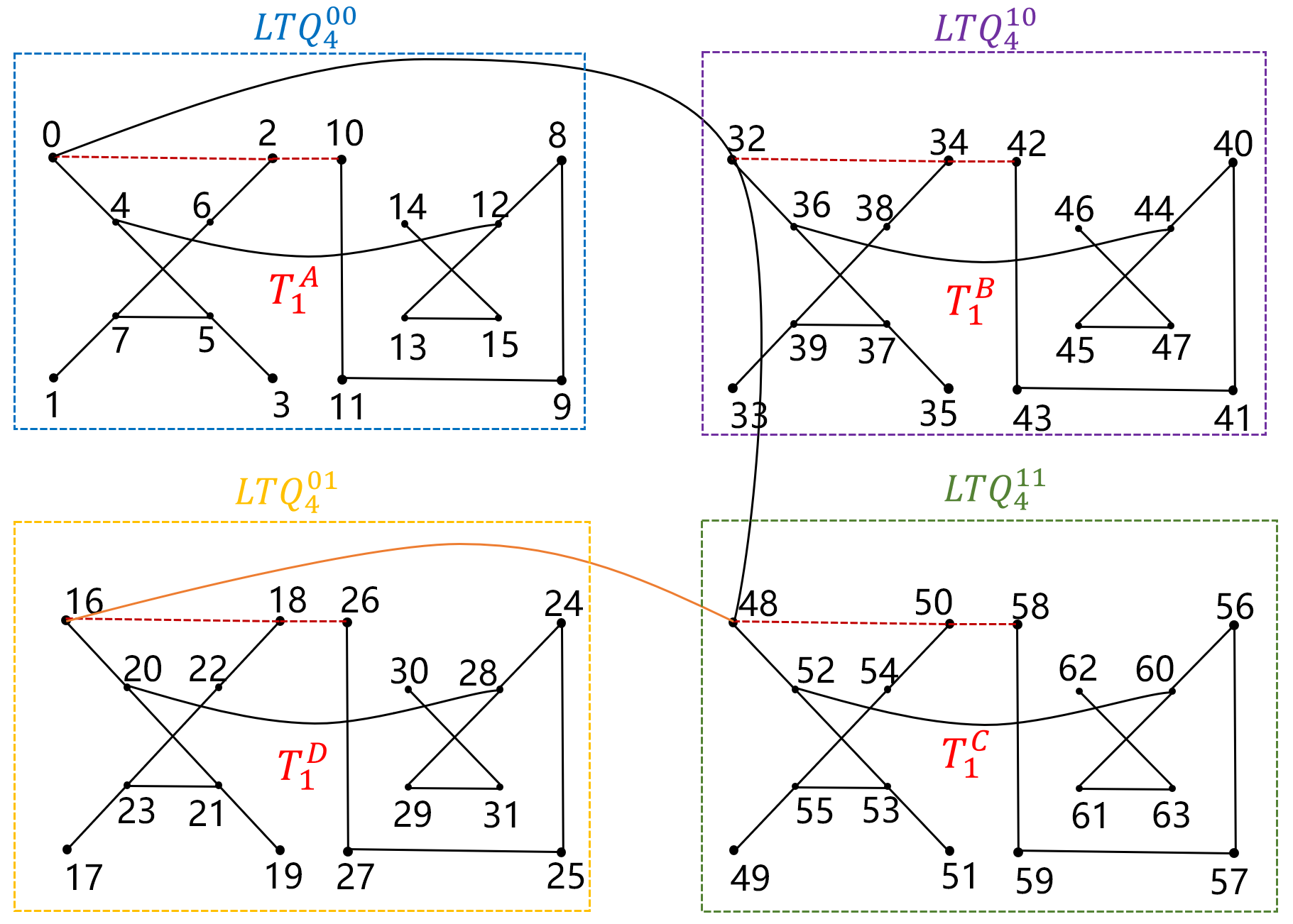}}
\subfigure[$T_2$]{
\label{LTQ6-T2}
\includegraphics[height=5.5cm]{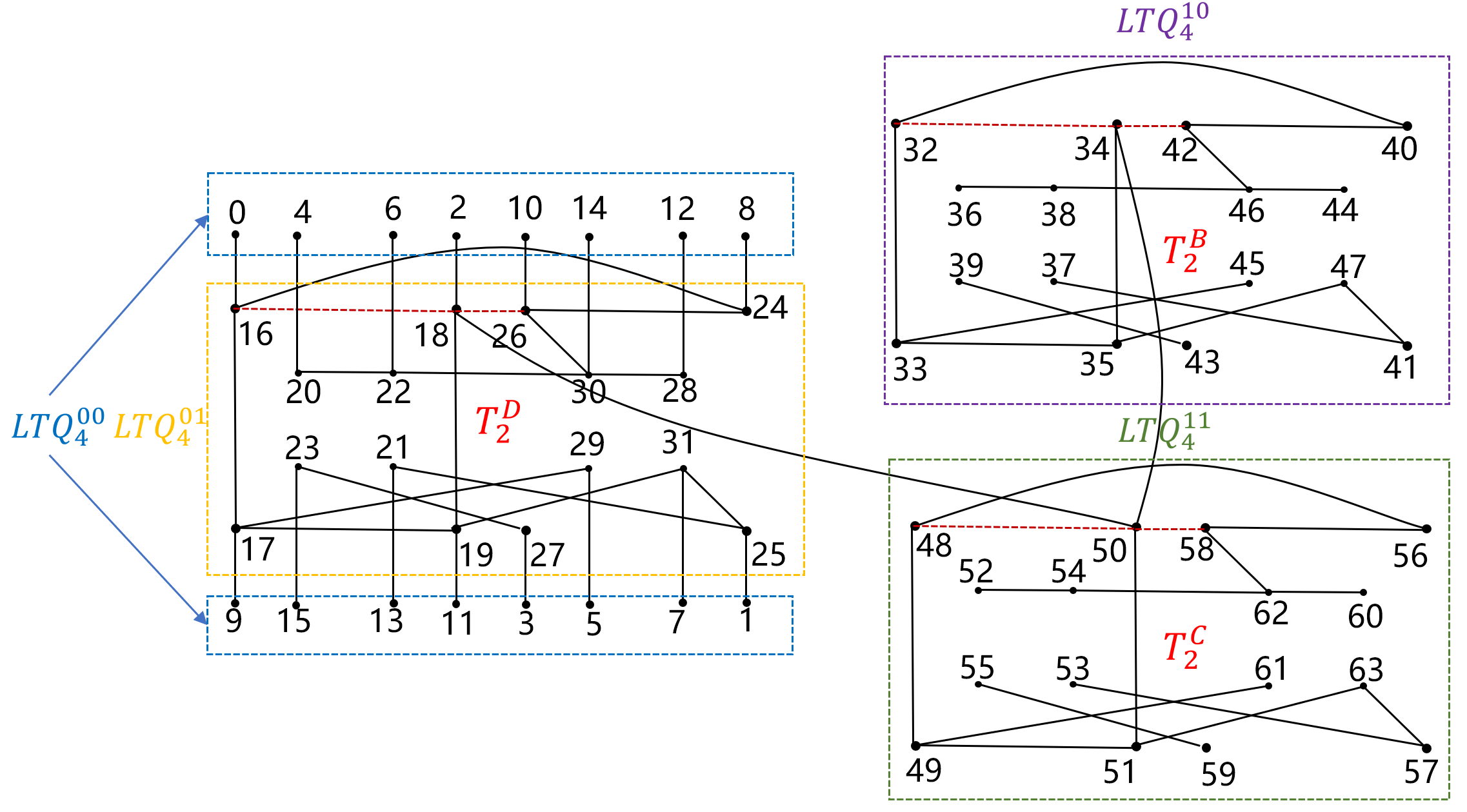}}
\subfigure[$T_3$]{
\label{LTQ6-T3}
\includegraphics[height=5.5cm]{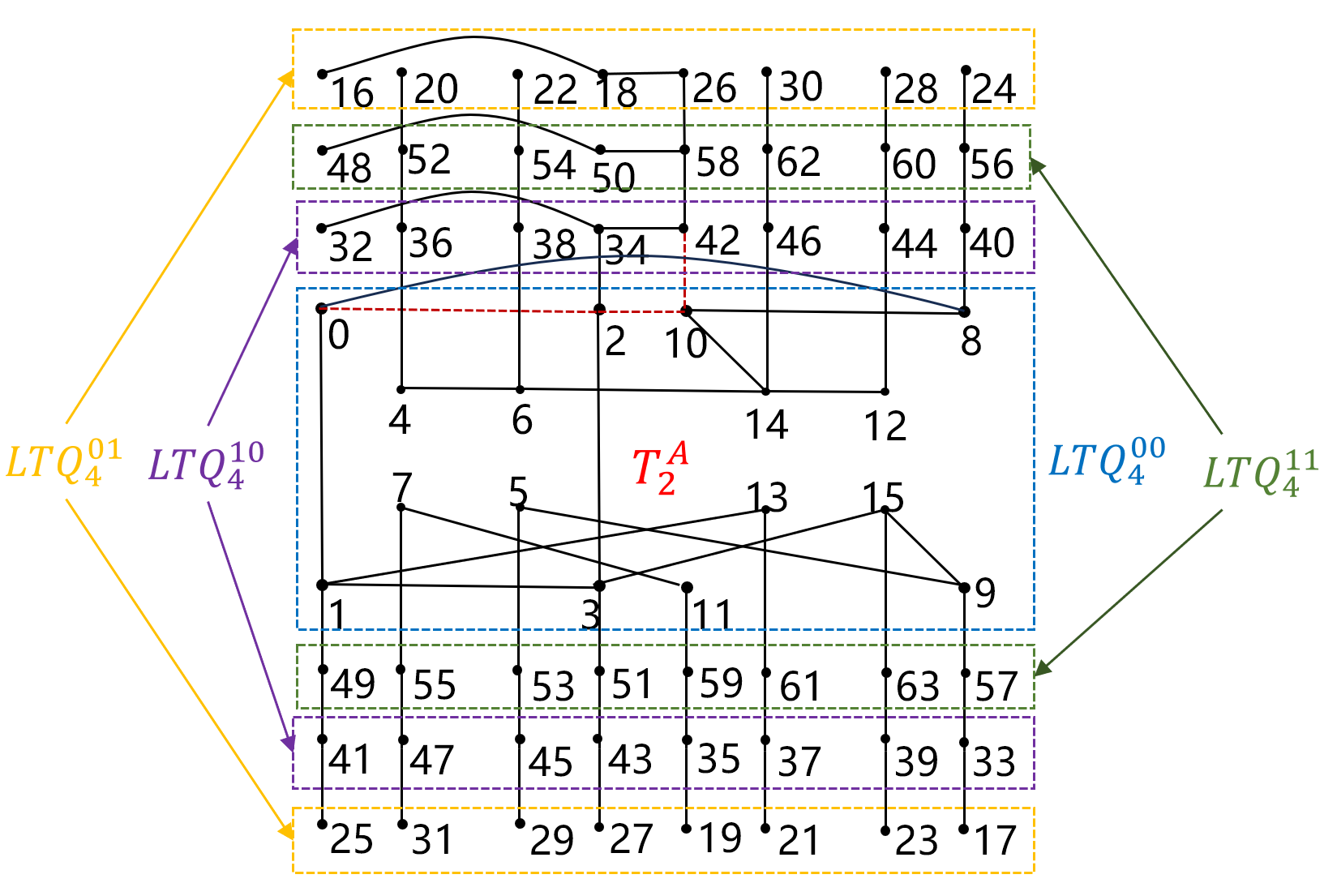}}
\caption{Three CEISTs $T_1$, $T_2$ and $T_3$ in $LTQ_6$.}
\label{fig6}
\end{figure}

\medskip
\noindent \textbf{Example 2.} By algorithm {\it Even\_CEISTs}, Fig.~5 demonstrates the construction of two CEISTs $T_1$ and $T_2$ in $LTQ_4$, and Fig.~6 demonstrates the construction of three CEISTs $T_1$, $T_2$ and $T_3$ in $LTQ_6$. We take $LTQ_6$ as an example, firstly, the two CEISTs $\gamma_A (4)= \{T_1^A, T_2^A\}$ and the path $P_A: 0$-$2$-$10$ in $LTQ_4$ as the input of the algorithm are presented in Fig.~5. Next, according to the step 1 of the algorithm {\it Even\_CEISTs}, $T_1^B, T_2^B$, $T_1^C, T_2^C$, $T_1^D, T_2^D$ and $P_B: 32$-$34$-$42$, $P_C: 48$-$50$-$58$, $P_D: 16$-$18$-$26$ are obtained. Then, according to the step 2 of the algorithm {\it Even\_CEISTs}, $T_1$ is constructed by connecting $T_1^A$, $T_1^B$, $T_1^C$ and $T_1^D$, through edges (0, 32), (32, 48) and (48, 16). $T_2$ is constructed by connecting $T_2^B$, $T_2^C$ and $T_2^D$ through edges (34, 50) and (50, 18), and all the edges between $A$ and $D$. $T_3$ is constructed by the following steps: (1) $T_2^A$ is contained in $T_3$. (2) First, connect all the even edges between $A$ and $B$ except (0, 32) and (10, 42), all the even edges between $B$ and $C$ except (32, 48) and (34, 50), and all the even edges between $C$ and $D$ except (16, 48) and (18, 50). (3) Then, connect all the odd edges between $A$ and $C$, all the odd edges between $C$ and $B$, and all the odd edges between $B$ and $D$. (4) Connect (32, 34), (34, 42), (48,50), (50, 58), (16, 18), (18, 26). Finally, the path $P$ is 0-2-10-42.

Then we synthesize the two algorithms and propose an integrated algorithm, named algorithm {\it CEISTs\_$LTQ$}, to generate $\lfloor \frac{n}{2} \rfloor$ CEISTs in $LTQ_n$, where $n \geq 2$.

\medskip
\begin{Verbatim}[commandchars=\\\{\},
framerule=0.6mm,frame=lines,codes={\catcode`$=3\catcode`^=7\catcode`_=8}]
\textbf{Algorithm {\it CEISTs\_$LTQ$}}
\textbf{Input:} Integer $n$ , with $n \geq 2$.
\textbf{Output:} $\gamma (n)$, for $n \geq 2$, and a path $P$: $v^1$-$v^2$-$\cdots$-$v^{\frac{n}{2}+1}$ when $n$ is even.
\textbf{Begin}
1: if $n$ = 2 then
2:     $V(T_1) = \{0, 1, 2, 3\}$, $E(T_1) = \{(0,1), (1,3), (3,2)\}$,
3:     $v^1 = 0, v^2 = 2$.
4: if $n \geq 3$ and $n$ is odd then
5:     $\gamma (n-1), P_A \gets {\it CEISTs\_LTQ}(n-1)$,
6:     $\gamma (n) \gets {\it Odd\_CEISTs} (\gamma (n-1), P_A)$.
7: if $n \geq 4$ and $n$ is even then
8:     $\gamma (n-2), P_A \gets {\it CEISTs\_LTQ}(n-2)$,
9:     $\gamma (n-1), P \gets {\it Even\_CEISTs}(\gamma (n-2), P_A)$.
10: return $\gamma (n)$, and path $P$ for $n$ is even.
end
\end{Verbatim}

\subsection{Correctness of {\it CEISTs\_$LTQ$}}

\begin{sloppypar}
 To verify the correctness of the CEISTs obtained by algorithm {\it CEISTs\_$LTQ$}, we present the following theorems.

 \begin{theorem} \rm For $n \geq 3$ and $n$ is odd, $T_1, T_2, ..., T_{\lfloor\frac{n}{2}\rfloor}$ constructed by algorithm {\it CEISTs\_$LTQ$} are $\lfloor\frac{n}{2}\rfloor$ CEISTs in $LTQ_n$.
\end{theorem}

\begin{proof}
By the step 1 of algorithm {\it Odd\_CEISTs}, $T_i^B$ is constructed by adding $2^{n-1}$ to each vertex in $T_i^A$, we have $ E(T_i^B) \cap E(T_j^B)= \emptyset$, for any $1 \leq i \textless j \leq \lfloor\frac{n}{2}\rfloor$, and $P_B$ is constructed by adding $2^{n-1}$ to each vertex in $P_A$, we have $V(P_A) \cap V(P_B) = \emptyset$. Since $|v_A^i|$ is even, for $1 \leq i \leq \lfloor\frac{n}{2}\rfloor$, according to Lemma 1, $v_A^i$ and $v_B^i$ are two adjacent vertices. Thus $(v_A^i, v_B^i) \neq (v_A^j, v_B^j)$, for any $1 \leq i \textless j \leq \lfloor\frac{n}{2}\rfloor$, $\{E(T_i^A) \cup E(T_i^B) \cup \{(v_A^i, v_B^i)\}\} \cap \{E(T_j^A) \cup E(T_j^B) \cup \{(v_A^j, v_B^j)\}\} = \emptyset$, for any $1 \leq i \textless j \leq \lfloor\frac{n}{2}\rfloor$. Therefore, $E(T_i) = \{E(T_i^A) \cup E(T_i^B) \cup \{(v_A^i, v_B^i)\}\}, E(T_i) \cap E(T_j)= \emptyset$, for any $1 \leq i \textless j \leq \lfloor\frac{n}{2}\rfloor$, there exist $\lfloor\frac{n}{2}\rfloor$ CEISTs $T_1, T_2,..., T_{\lfloor\frac{n}{2}\rfloor}$ in $LTQ_n$, where $n \geq 3$ and $n$ is odd.
\end{proof}

 \begin{theorem}\rm For $n \geq 4$ and $n$ is even, $T_1,T_2,...,T_{\frac{n}{2}}$ constructed by algorithm {\it CEISTs\_$LTQ$} are $\frac{n}{2}$ CEISTs in $LTQ_n$.
\end{theorem}

 \begin{proof}
 By the step 1 of algorithm {\it Even\_CEISTs}, trees $T_i^B$, $T_i^C$ and $T_i^D$ are constructed by adding $2^{n-1}$, $3*2^{n-1}$ and $2^{n-2}$ to each vertex in $T_i^A$, respectively, we have $\{ E(T_i^r) \cap E(T_j^r)= \emptyset | r \in \{B, C, D\}, 1 \leq i \textless j \leq \frac{n}{2}-2 \}$, and $P_B$, $P_C$, $P_D $ are constructed by adding $2^{n-1}, 3*2^{n-1}, 2^{n-2}$ to each vertex in $P_A$, respectively, we have $V(P_A) \cap V(P_B) \cap V(P_C) \cap V(P_D)= \emptyset$. Since $|v_A^i|$ is even, for $i \in \{1,2,..., \frac{n}{2}\}$, according to Lemma \ref{lem1}, $v_A^i$ and $v_B^i$ are two adjacent vertices, $v_B^i$ and $v_C^i$ are two adjacent vertices, $v_C^i$ and $v_D^i$ are two adjacent vertices, and $v_A^i$ and $v_D^i$ are two adjacent vertices. We have the following cases:

\medskip
 \textbf{Case 1.} Construct $T_i$, for $1 \leq i \leq \frac{n}{2}-2$. By the step 2 of algorithm {\it Even\_CEISTs}, we can know $E(T_i)=E(T_i^A) \cup E(T_i^B) \cup E(T_i^C) \cup E(T_i^D) \cup \{(v_A^i, v_B^i), (v_B^i, v_C^i), (v_C^i, v_D^i)\}$. Therefore, $E(T_i) \cap E(T_j)= \emptyset$, for any $1 \leq i \textless j \leq \frac{n}{2}-2$.

\medskip
 \textbf{Case 2.} Construct $T_{\frac{n}{2}-1}$. From the topology of $LTQ_n$, We know both even vertices and odd vertices are adjacent only between $A$ and $D$, or between $B$ and $C$, the simple topology of $LTQ_n$ is presented in Fig. \ref{fig7}. Thus, we choose $T_{\frac{n}{2}-1}^B, T_{\frac{n}{2}-1}^C$ and $T_{\frac{n}{2}-1}^D$ as the infrastructure, and connect them by $(v_B^{\frac{n}{2}-1}, v_C^{\frac{n}{2}-1}), (v_C^{\frac{n}{2}-1}, v_D^{\frac{n}{2}-1})$. Then, we connect all the vertices between $A$ and $D$ to obtain tree $T_{\frac{n}{2}-1}$. Obviously, $E(T_i) \cap E(T_j)= \emptyset$, for any $1 \leq i \textless j \leq \frac{n}{2}-1$.

\begin{figure}[htb]
\centering
\includegraphics[height=4cm]{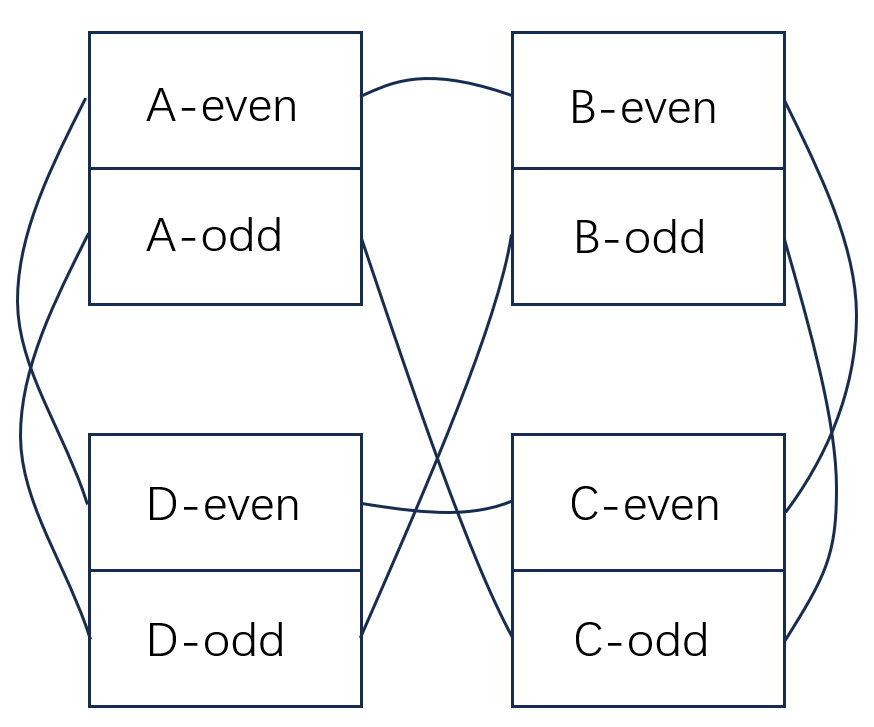}
\caption{The topology of $LTQ_n$.} \label{fig7}
\end{figure}

\textbf{Case 3.} Construct $T_{\frac{n}{2}}$. By Case 1 and Case 2, we know the unused edges are $E(T_{\frac{n}{2}-1}^A)$, even edges between $A$ and $B$ except for $(v_A^i, v_B^i) (1 \leq i \leq {\frac{n}{2}-2})$ and $(v_A^{\frac{n}{2}}, v_B^{\frac{n}{2}})$, even edges between $B$ and $C$ except for $(v_B^i, v_C^i) (1 \leq i \leq {\frac{n}{2}-1})$, even edges between $C$ and $D$ except for $(v_C^i, v_D^i) (1 \leq i \leq {\frac{n}{2}-1})$, all the odd edges in $LTQ_n$, and $(v_r^j, v_r^{j+1})$, for $r \in \{B, C, D\}$, and $j = 1$ to $\frac{n}{2}-1$. Thus, we choose $T_{\frac{n}{2}-1}^A$ as the infrastructure, (1) Connect all the even edges between $A$ and $B$ except for $(v_A^i, v_B^i)$ and $(v_A^{\frac{n}{2}}, v_B^{\frac{n}{2}})$ , for $1 \leq i \leq {\frac{n}{2}-2}$. Then, we connect $v_B^i(1 \leq i \leq {\frac{n}{2}-2})$ and $v_B^{\frac{n}{2}}$ to the tree by $v_B^{\frac{n}{2}-1}$, that is we connect $(v_B^i, v_B^{i+1}) (1 \leq i \leq {\frac{n}{2}-1})$. (2) Connect all the even edges between $B$ and $C$ except for $(v_B^i, v_C^i) (1 \leq i \leq {\frac{n}{2}-1})$. Then we connect $v_C^i (1 \leq i \leq {\frac{n}{2}-1})$ to the tree by $v_C^{\frac{n}{2}}$, that is we connect $(v_C^i, v_C^{i+1}) (1 \leq i \leq {\frac{n}{2}-1})$. (3) Connect all the even edges between $C$ and $D$ except for $(v_C^i, v_D^i) (1 \leq i \leq {\frac{n}{2}-1})$. Then we connect $v_D^i (1 \leq i \leq {\frac{n}{2}-1})$ to the tree by $v_D^{\frac{n}{2}}$, that is we connect $(v_D^i, v_D^{i+1}) (1 \leq i \leq {\frac{n}{2}-1})$. (4) Connect all the odd edges in $LTQ_n$. Thus, For $n \geq 4$ and $n$ is even, $T_1,T_2,...,T_{\frac{n}{2}}$ constructed by algorithm {\it CEISTs\_$LTQ$} are $\frac{n}{2}$ CEISTs in $LTQ_n$.
\end{proof}

\begin{theorem} \rm For $n \geq 2$ and $n$ is even, after $\frac{n}{2}$ CEISTs are constructed, there still remain $\frac{n}{2}$ unused edges.
\end{theorem}

 \begin{proof}
  Since $LTQ_n$ has $n\cdot2^{n-1}$ edges, and each spanning tree in $LTQ_n$ has $2^n-1$ edges, $\frac{n}{2}$ spanning trees have $(2^n-1)\cdot\frac{n}{2} = n\cdot2^{n-1} - \frac{n}{2}$ edges. Thus, after $\frac{n}{2}$ CEISTs are constructed, there still remain $\frac{n}{2}$ unused edges. In order to reduce the time complexity, we specify the $\frac{n}{2}$ unused edges are $v_A^{1}$-$v_A^{2}$-$\cdots-v_A^{\frac{n}{2}}$-$v_A^{\frac{n}{2}+1}$, where $|v_A^{1}| = 0, |v_A^{\frac{n}{2}+1}|= |v_A^{\frac{n}{2}}|+2^{n-1}$.

In summary, for $n \geq 2$, $T_1, T_2,..., T_{\frac{n}{2}}$ constructed by algorithm {\it CEISTs\_$LTQ$} are $\frac{n}{2}$ CEISTs in $LTQ_n$.
\end{proof}

\begin{theorem}\label{Th4} \rm For integer $n \geq 2$, algorithm {\it CEISTs\_$LTQ$} obtains $\lfloor\frac{n}{2}\rfloor$ CEISTs in
 $O(n \cdot 2^n)$   time, where $n$ is the  dimension of $LTQ_n$.
\end{theorem}

 \begin{proof} Let $T(n)$ denote the running time of the algorithm {\it CEISTs\_$LTQ$}. The time complexity of algorithm 1 and algorithm 2 is $O({2^n})$, we have a recurrence equation that bounds:
$$T(n) = \begin{cases}
1, n = 2; \\
T(n-1) + O (2^n), n \text{ is odd;}  \\
T(n-2) + O (2^n), n \text{ is even.} \\
\end{cases}$$
Solving the recurrence equation yields that the time complexity is $O(n\cdot{2^n})$.
\end{proof}
\end{sloppypar}

\subsection{CEISTs in $LTQ$ vs. in hypercubes/crossed cubes}

 We make some remarks pertaining to CEISTs algorithms for hypercubes and crossed cubes.
As has been noted, algorithms for embedding CEISTs in hypercubes and crossed cubes have been provided in [8] and [18], respectively.
Although {\it CEISTs\_$LTQ$} also uses a recursive scheme, it is not a straightforward, direct application of the existing methods in [8] or [18]. Due to these cubes' differences in topology, different techniques are used for the task.

Note that when $n$ is odd, $LTQ_n$ and $LTQ_{n-1}$ have {\it the same number} (i.e. $\frac{n-1}{2}$) of CEISTs. That means we can just splice the CEISTs in the two $LTQ_{n-1}$s to build the $\frac{n-1}{2}$ CEISTs in $LTQ_n$. That is, we just need to choose splicing edges between the two $LTQ_{n-1}$s, and no additional CEISTs need to be found in the process. For this case, $LTQ$/hypercube/crossed cube's treatments are similar.

\medskip
The complication arises when $n$ is even. Now the $LTQ_n$'s $\frac{n}{2}$ CEISTs are recursively constructed from four $LTQ_{n-2}$s, each of which contains $\frac{n-2}{2}$ CEISTs:
\begin{itemize}
\item[]{\bf Step $1$}: Since $n-2$ is even, four $LTQ_{n-2}$s can be spliced into one $LTQ_n$, which still contains $\frac{n-2}{2}$ CEISTs;

\item[]{\bf Step $2$}: Choose unused edges inside and between the four $LTQ_{n-2}$s to build one more CEIST, so that we have $\frac{n-2}{2}+1=\frac{n}{2}$ CEISTs.
\end{itemize}

It is in Step 2 above that the method for $LTQ$ is more complex than in hypercubes/crossed cubes.
Due to the less-regular, ``locally twisted'' connection between $LTQ$ subcubes, the edge selection techniques for hypercubes/crossed cubes fail to work. A more restricted selection procedure is carried out by {\it CEISTs\_$LTQ$}.

\medskip
As a matter of fact, the edge selection procedure for {\it CEISTs\_$LTQ$} will also work for both hypercubes and crossed cubes. However, it would introduce restrictions that are unnecessary when selecting CEISTs edges in hypercubes and crossed cubes.

\section{Implementation and simulation}

\begin{figure}[!h]
\centering
\subfigure[$T_1$]{
\label{7-T1}
\includegraphics[height=5.2cm]{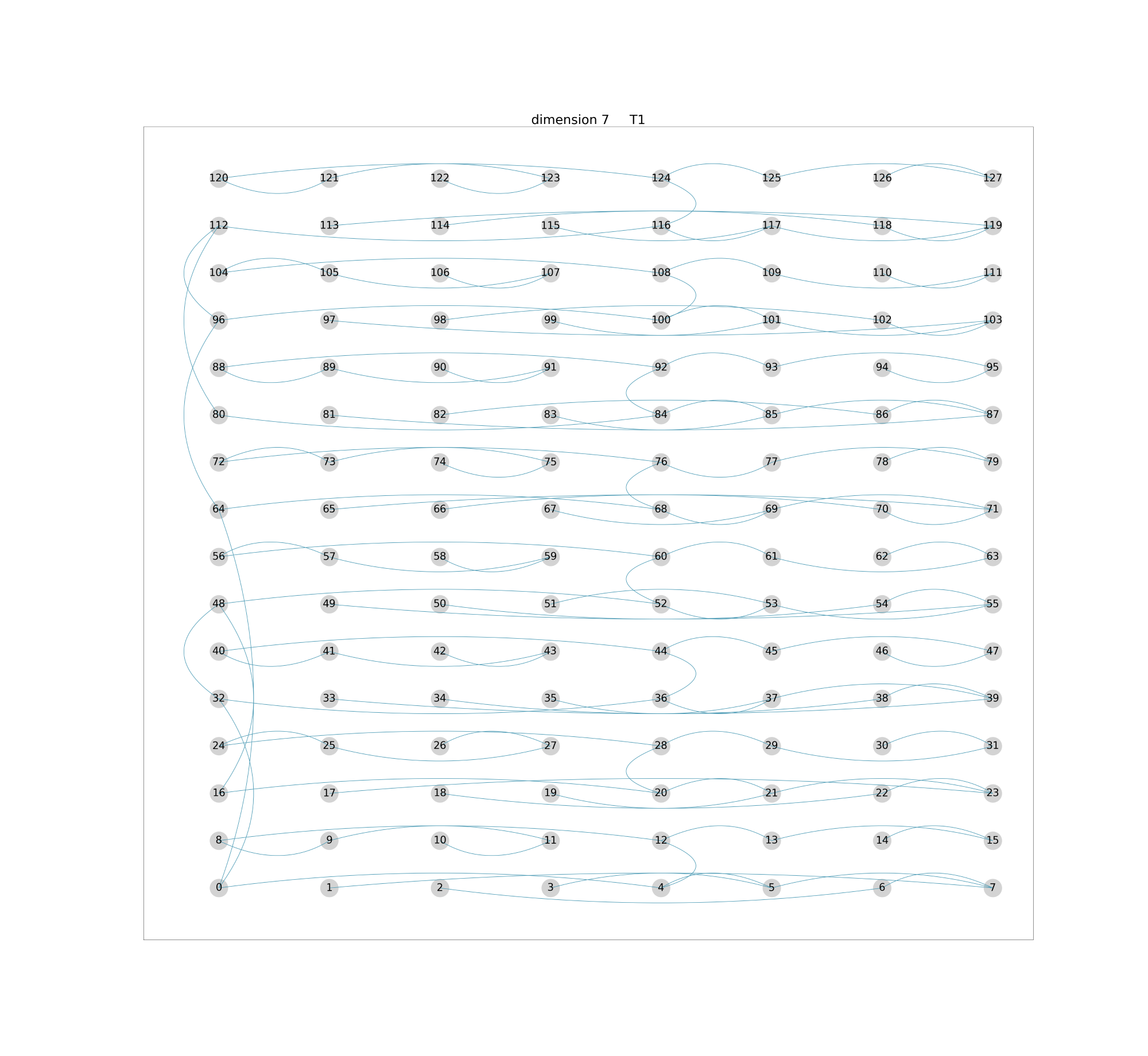}}
\subfigure[$T_2$]{
\label{7-T2}
\includegraphics[height=5.2cm]{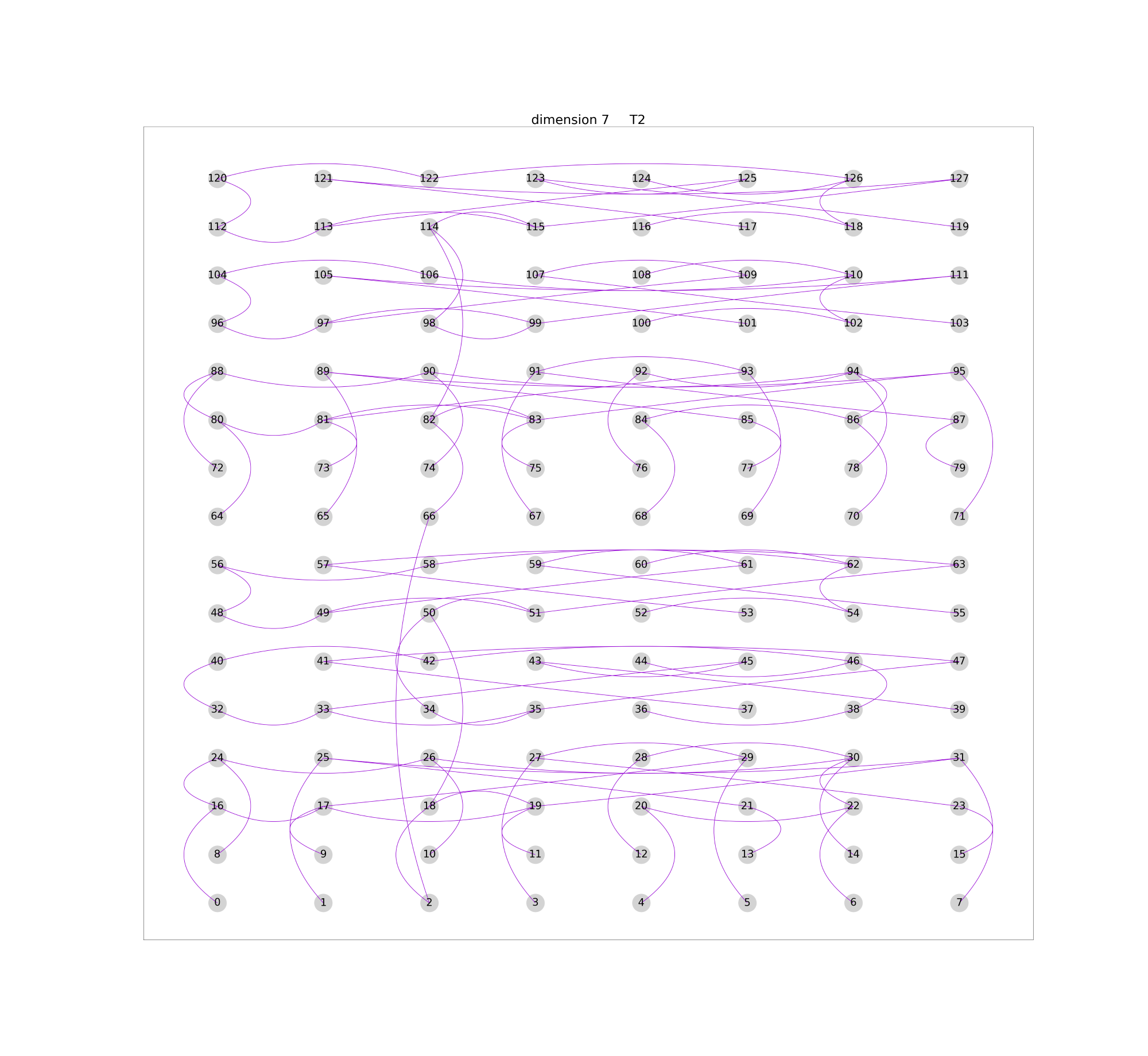}}
\subfigure[$T_3$]{
\label{7-T3}
\includegraphics[height=5cm]{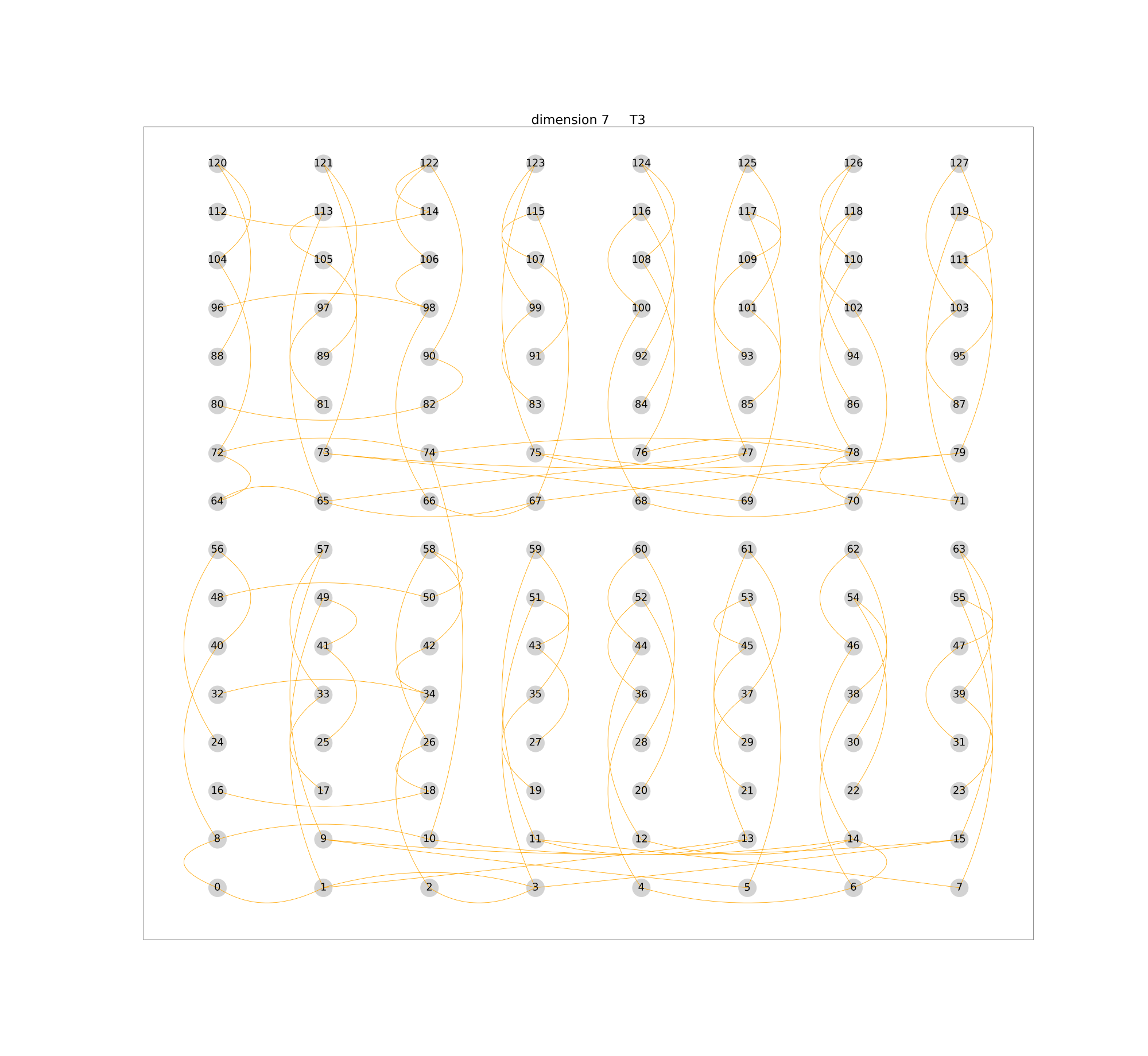}}
\caption{Three CEISTs in $LTQ_7$.}
\label{fig8}
\end{figure}

\begin{figure}[!h]
\centering
\subfigure[$T_1$]{
\label{8-T1}
\includegraphics[height=5.2cm]{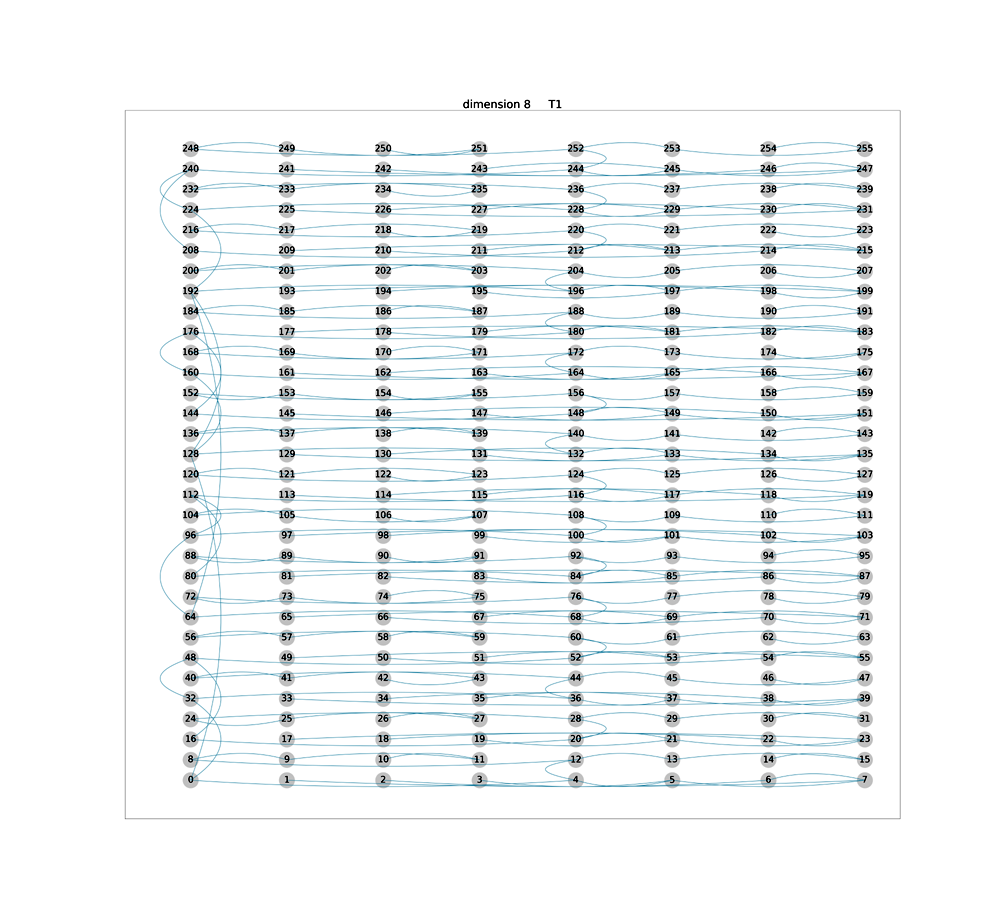}}
\hspace{0.01cm}
\subfigure[$T_2$]{
\label{8-T2}
\includegraphics[height=5.2cm]{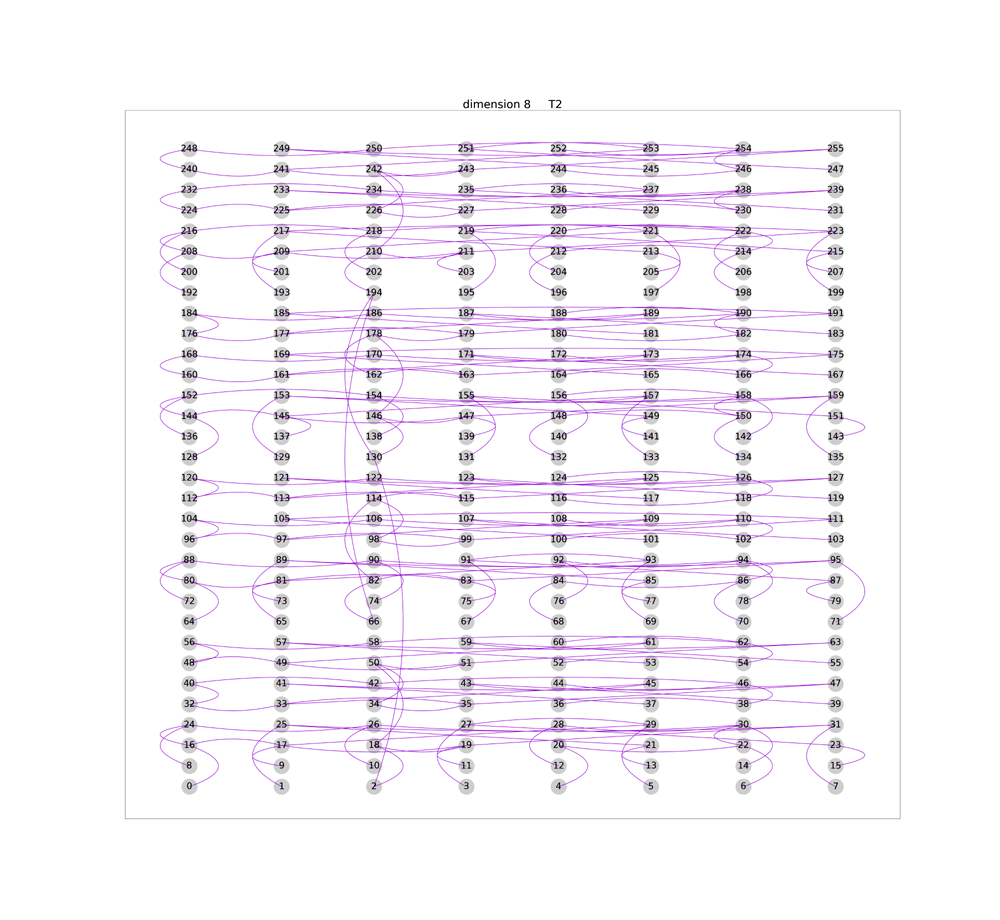}}
\hspace{0.01cm}
\subfigure[$T_3$]{
\label{8-T3}
\includegraphics[height=5.2cm]{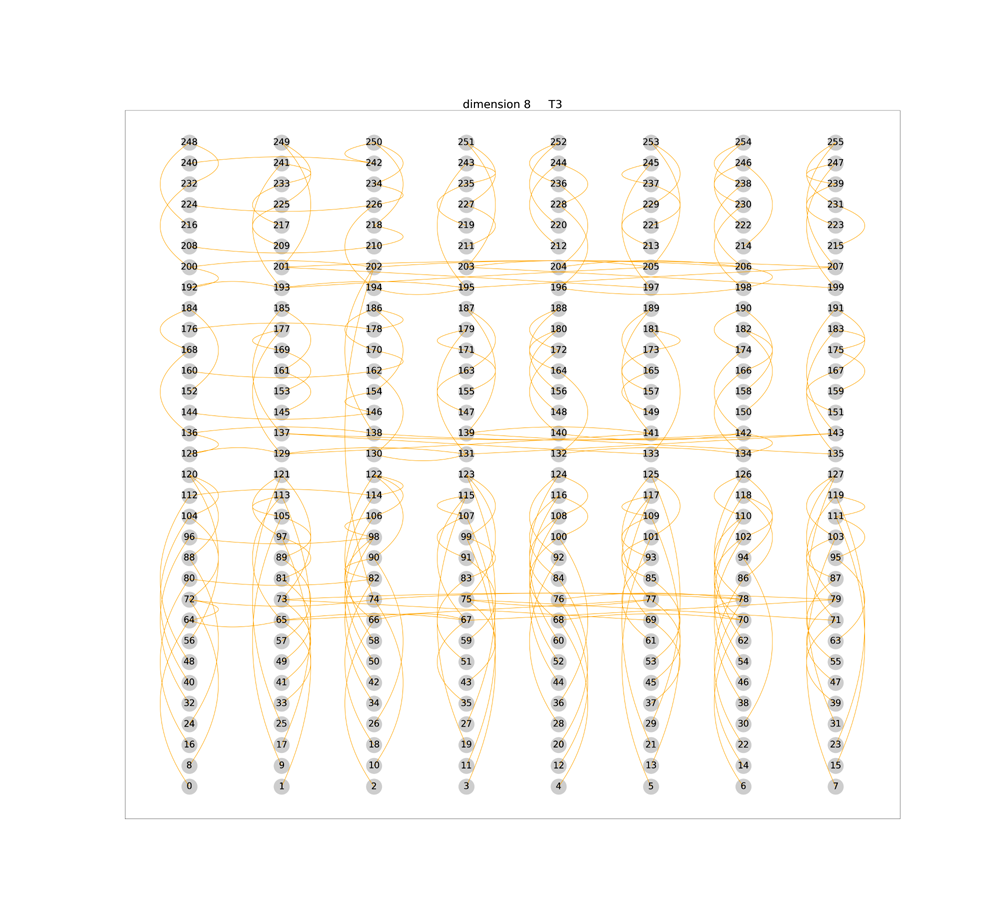}}
\hspace{0.01cm}
\subfigure[$T_4$]{
\label{8-T4}
\includegraphics[height=5.2cm]{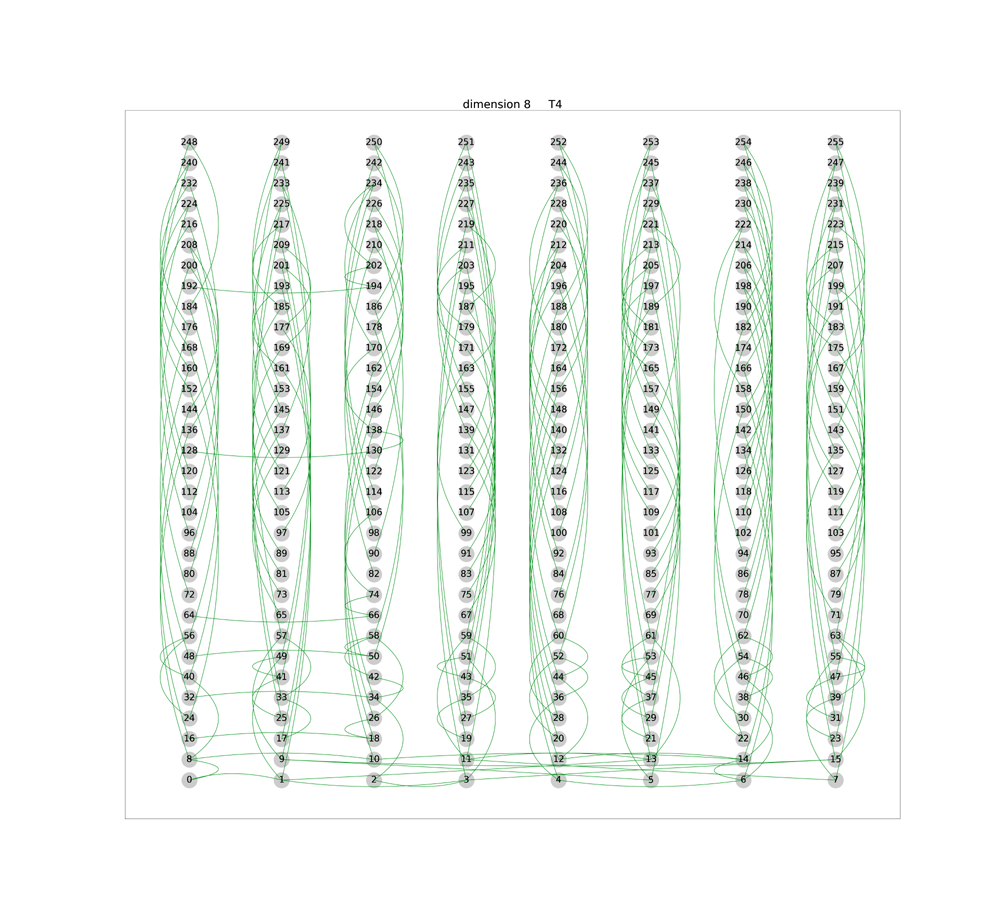}}
\caption{Four CEISTs in $LTQ_8$.}
\label{fig9}
\end{figure}

     To attest algorithm {\it CEISTs\_$LTQ$}'s validity, it was actually implemented using the programming language Python. The program's main methods exactly follow the steps outlined in {\it CEISTs\_$LTQ$}. Fig.~\ref{fig8} illustrates the algorithm output, CEISTs in an odd $LTQ_7$, while Fig.~\ref{fig9} illustrates the CEISTs in an even $LTQ_8$.


\medskip
CEISTs can be used for efficient broadcasting, reliable broadcasting and secure distribution of information. Then, we can divide a message into several
data packets, encrypt every packet, and transmit them through multiple CEISTs, finally we decrypt and merge them when we receive all the packets to achieve efficient broadcasting and secure distribution of information.

 We simulate  the scenario where a message of size less than $1 M$ is broadcast from a source node to all nodes
in $LTQ_n$, using multiple CEISTs. The most common Ethernet frame length can carry about 1500 bytes of the message (excluding the initial preamble, frame delimiter, and the frame check sequence at the end) [23]. Thus, the message can be divided into $1 M /1500$ $bytes = 700$ (data packets). We compare $\lfloor \frac{n}{2} \rfloor$ CEISTs as transmission channels with a single spanning tree as transmission channel for data broadcasting. We employ a round-robin strategy to call $\lfloor \frac{n}{2} \rfloor$ channels for packet transmission to balance the load of all transmission channels. That is the first $\lfloor \frac{n}{2} \rfloor$ packets are transmitted by the $\lfloor \frac{n}{2} \rfloor$ channels in sequence, the $\lfloor \frac{n}{2} \rfloor + 1$ packet is transmitted by the first channel, the $\lfloor \frac{n}{2} \rfloor + 2$ packet is transmitted by the second channel, and so on.

\medskip
We use the following two metrics to evaluate broadcasting efficiency, one is the \emph{maximum broadcasting latency} (MBL for short), and the other is the \emph{average broadcasting latency} (ABL for short). For $ 1 \leq k \leq \lfloor \frac{n}{2} \rfloor$, let $mt(k)$ be the delivery the whole message time between the farthest two vertices in the $kth$ tree. There are $s = 2^{2n-1}-2^{n-1}$ pairs of vertices in $LTQ_n$. For $ 1 \leq i \leq s$, let $arbt(i)$ be the maximum distance between the $ith$ pairs of vertices in $\lfloor \frac{n}{2} \rfloor$ CEISTs, and let $t(i)$ be the distance between the $ith$ pairs of vertices in one spanning tree.

\begin{figure}[!h]
\centering
\includegraphics[height=5.2cm]{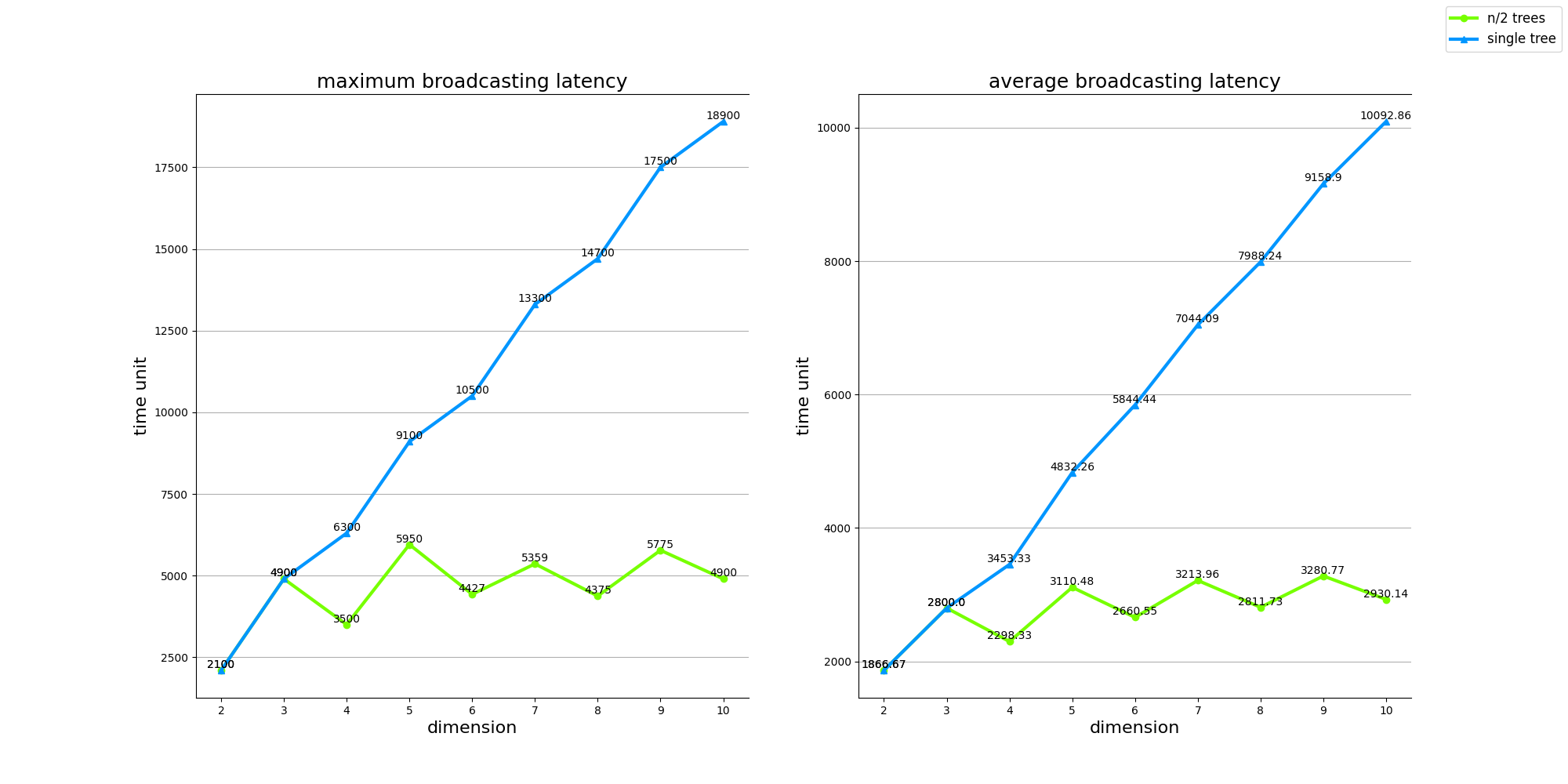}
\caption{The comparisons of ABL and MBL between $\lfloor \frac{n}{2} \rfloor$ CEISTs and a single spanning tree.} \label{fig10}
\end{figure}

\medskip
Suppose a message is divided into $x$ data packets. Firstly, the MBL of broadcasting latency using $\lfloor \frac{n}{2} \rfloor$ CEISTs is: $MBL =\lceil \frac{x}{\lfloor\frac{n}{2} \rfloor} \rceil \cdot \mathop{\max}\limits_{1 \leq k \leq {\lfloor \frac{n}{2} \rfloor}} \{ mt(k) \}$, and the MBL of broadcasting latency using a single spanning tree is: $MBL = x  \cdot \mathop{\min}\limits_{1 \leq k \leq {\lfloor \frac{n}{2} \rfloor}} \{ mt(k) \}$. Secondly, the ABL of broadcasting latency using $\lfloor \frac{n}{2} \rfloor$ CEISTs is: $ABL = \frac{\mathop{\sum}\limits_{i=1}^s arbt(i)}{s}$, and the ABL of broadcasting latency using a single spanning tree is: $ABL = \frac{\mathop{\sum}\limits_{i=1}^s t(i)}{s}$. All the experimental results showing ABL and MBL are depicted in Fig. \ref{fig10}. To demonstrate the efficiency of broadcasting in multiple CEISTs, we choose the spanning tree with the minimal $mt(k)$ in $\lfloor\frac{n}{2} \rfloor$ CEISTs to calculate the MBL of the single spanning tree, and choose the spanning tree with the minimal ABL in $\lfloor\frac{n}{2} \rfloor$ CEISTs to calculate the ABL of the single spanning tree.

For both ABL and MBL, the performance of broadcasting latency using $\lfloor \frac{n}{2} \rfloor$ CEISTs is better than that using a single spanning tree. As the network size grows, in the odd dimension and even dimension, the ABL and MBL of $\lfloor \frac{n}{2} \rfloor$ CEISTs will rise, respectively, but the growth rate is very slow compared with the single spanning tree.

\section{Conclusion}

     We proposed, and proved correctness for, an $O(n \cdot 2^n)$  algorithm, named {\it CEISTs\_$LTQ$}, to construct $\lfloor\frac{n}{2}\rfloor$ CEISTs in the {\it locally twisted cube} network $LTQ_n$, where $n \geq 2$ is the dimension. The number of CEISTs constructed by our algorithm is optimal. Experiments were conducted to verify the validity of our algorithm, and to simulate broadcasting $LTQ_n$ using the CEISTs. It is worth pointing out that our proposed algorithm for $LTQ_n$ can also be used to construct CEISTs in hypercube $Q_n$ and crossed cube $CQ_n$, which makes it a more general algorithm.

Many directions can be pursued for continuing our work. Among them, for example, CEIST-embedding in the presence of faulty (therefore missing) nodes can be explored [25, 26]. Since $\lfloor \frac{n}{2} \rfloor$ CEISTs have already been constructed for $Q_n$, $CQ_n$, and now for locally twisted cube $LTQ_n$, a rather reasonable conjecture would be that all hypercube variants, or even all bijective connection networks, have $\lfloor \frac{n}{2} \rfloor$ CEISTs.

\section*{Acknowledgment}

     This work was supported by the National Natural Science Foundation of China (Nos. 62272333, 62172291, U1905211) and Jiangsu Province Department of Education Future Network Research Fund Project (FNSRFP-2021-YB-39).

\end{document}